\documentclass[twocolumn]{aastex62}
\usepackage{amsmath,amstext}
\usepackage[T1]{fontenc}
\usepackage{apjfonts} 
\usepackage[figure,figure*]{hypcap}
\usepackage{verbatim}
\usepackage{graphicx}

\shorttitle{MOJAVE XVI}
\shortauthors{Hodge et al.}

\begin{document}

\title{MOJAVE XVI: Multi-Epoch Linear Polarization Properties of Parsec-Scale AGN Jet Cores}

\author{M. A. Hodge}
\affiliation{Department of Physics and Astronomy, Purdue University, 525 Northwestern Avenue,
West Lafayette, IN 47907, USA;
\email{hodge2@purdue.edu; mlister@purdue.edu}}
\author{M. L. Lister}
\affiliation{Department of Physics and Astronomy, Purdue University, 525 Northwestern Avenue,
West Lafayette, IN 47907, USA;
\email{hodge2@purdue.edu; mlister@purdue.edu}}
\author{M. F. Aller}
\affiliation{Department of Astronomy, University of Michigan, 311 West Hall, 1085 S. University Avenue,
 Ann Arbor, MI 48109, USA;}
\author{H. D. Aller}
\affiliation{Department of Astronomy, University of Michigan, 311 West Hall, 1085 S. University Avenue,
 Ann Arbor, MI 48109, USA;}
\author{Y. Y. Kovalev}
\affiliation{Astro Space Center of Lebedev Physical Institute, Profsoyuznaya 84/32, 117997 Moscow, Russia;}
\affiliation{Moscow Institute of Physics and Technology, Dolgoprudny, Institutsky per., 9, Moscow region, 141700, Russia}
\affiliation{Max-Planck-Institut f{\"u}r Radioastronomie, Auf dem H{\"u}gel 69, D-53121 Bonn, Germany;}
\author{A. B. Pushkarev}
\affiliation{Crimean Astrophysical Observatory, 298409 Nauchny, Crimea, Russia;}
\affiliation{Astro Space Center of Lebedev Physical Institute, Profsoyuznaya 84/32, 117997 Moscow, Russia;}
\author{T. Savolainen}
\affiliation{Aalto University Department of Electronic and Nanoengineering, PL15500, FI-00076 Aalto, Finland;}
\affiliation{Aalto University Mets{\"a}hovi Radio Observatory, Mets{\"a}hovintie 114, FI-02540 Kylm{\"a}l{\"a}, Finland;}
\affiliation{Max-Planck-Institut f{\"u}r Radioastronomie, Auf dem H{\"u}gel 69, D-53121 Bonn, Germany;}

\begin{abstract}

We present an analysis of the core linear polarization properties of 387 parsec-scale active galactic nuclei (AGN) jets. Using 15 GHz VLBA data, we revisit the conclusions of the first paper in this series with multi-epoch measurements and more detailed analysis of a larger AGN sample which spans a broader range of synchrotron peak frequencies. Each AGN has been observed for at least five epochs between 1996 and 2017. We find that BL Lac objects have  core electric vector position angles (EVPAs) which tend towards alignment with the local jet direction; compared to flat spectrum radio quasars (FSRQs), their EVPAs are also less variable over time. The AGN cores which are most fractionally polarized and least variable in polarization have EVPAs that are closely aligned with the local jet direction; they also have low variability in EVPA. These results support the popular model of a standing transverse shock at the base of the jet which collimates the jet magnetic field perpendicular to the jet direction, increasing the fractional polarization and leading to greater polarization stability over time. High-synchrotron-peaked (HSP) BL Lac objects form a low luminosity, low fractional polarization population. The five narrow-line Seyfert 1 galaxies in our sample have low fractional polarization and large EVPA-jet misalignments. Although AGN detected at $\gamma$-rays are thought to be more Doppler boosted than non-detected AGN, we find no significant differences in fractional polarization based on detection by \(Fermi\)-LAT; the$\gamma$-loud AGN are, however, more variable in core EVPAs.

\end{abstract}

\keywords{BL Lacertae objects: general --- galaxies: active --- galaxies: jets --- polarization --- quasars: general --- radio continuum: galaxies}

\section{Introduction}

In the first paper in this series, we published a VLBA 15 GHz study of the parsec-scale linear polarization properties of 133 jets associated with compact radio loud active galactic nuclei \citep{MOJAVE_I}. A wealth of information has subsequently become available, but relatively few radio surveys have focused on the multi-epoch polarization properties of large numbers of AGN. Polarization is a useful tool for determining the magnetic field structure of the jet, and may be particularly important in assessing the inherent differences between subcategories of blazars. As jet emission can be highly variable---down to minute timescales in some bands---multi-epoch measurements can show drastic changes.

In the case of synchrotron radiation, the fractional polarization can theoretically reach as high as 70\% of the total intensity flux \citep{2011hea..book.....L}. The observed amount and characteristics are highly dependent on viewing angle, due to relativistic boosting and aberration \citep{2005MNRAS.356..859P, 2005MNRAS.360..869L}. According to radio unification theory, blazars and radio galaxies are the same objects, with blazars oriented at a closer angle to the line of sight. Furthermore, it is generally believed that the blazar optical subclasses, BL Lacs and flat spectrum radio quasars (FSRQs), are linked to FR-type I and II galaxies, respectively \citep{UP95}. Correspondingly, BL Lacs and FSRQs have been found to have noticeably different polarization properties \citep{1993ApJ...416..519C, 2003ApJ...589..733P, MOJAVE_I}. Possible factors that could cause this dissimilarity include: jet speeds and shock strength, leading to differing magnetic field order \citep{MOJAVE_I}; viewing angle \citep{2014AA...566A..59A}; the rate of emerging new jet components \citep{2000MNRAS.319.1109G};
and the amount of Faraday depolarization \citep{2000ApJ...533...95T, 2017MNRAS.467...83K}.
In addition, these polarization properties have been found to be dependent on synchrotron SED peak frequency and association with a $\gamma$-ray source \citep{2011ApJ...742...27L, 2016MNRAS.463.3365A, 2010IJMPD..19..943H}---characteristics which are not equally spread between optical subclasses.

This paper focuses on the parsec-scale radio cores of AGN \citep{1997ARA&A..35..607Z}, which are generally associated with the bright, unresolved, optically thick feature near the base of the jet in VLBA images. The polarization properties of the downstream jet emission will be presented in a future paper in this series. The cores typically exhibit a lower fractional polarization (\(m= (Q^2 + U^2)^{1/2}/I \)), where $I$, $Q$, and $U$ are intensities in each Stokes parameter) than the downstream jet \citep{MOJAVE_I}. Selection effects may be a partial reason for this finding \citep{2013EPJWC..6106001W}, but it is mainly thought to be a result of magnetic fields becoming more ordered with distance down the jet. In this respect, the electric vector position angle (EVPA=\(0.5 \arctan(U Q^{-1})\)) plays an important role in analysis; depending on the optical thickness and speed of the relativistic plasma within the jet \citep{2005MNRAS.360..869L}, the EVPA indicates the direction and orderedness of the magnetic field. Transverse shocks within the jet can order the magnetic field and orient the EVPA parallel to the jet direction \citep{1980MNRAS.193..439L}, while the emergence of bright moving jet features with arbitrary magnetic field direction at the core may rotate the EVPA \citep{cohensubmit} and affect the observed polarization fraction. In addition, the EVPA can be an essential indicator of the underlying magnetic field structure, either toroidal or helical \citep{2003NewAR..47..599G}. In order to explore these aspects of jet physics, we have carried out a multi-epoch linear polarization study of the MOJAVE AGN sample with observations gathered during the last three decades with the VLBA. We adopt a cosmology with $\Omega_{m}$ = 0.27, $\Omega_{\Lambda}$ = 0.73, and $H_{0}$ = 71 km s$^{-1}$ Mpc$^{-1}$ \citep{2009ApJS..180..330K}. All position angles are quoted in degrees east of north.

\section{AGN Sample and Observational Data} \label{sampledata}

In this paper we analyze the parsec-scale radio core polarization and total intensity properties of 387 AGN selected from the MOJAVE survey which have at least five VLBA epochs at 15 GHz between 1996 January 19 and 2016 December 31. These AGN all have $\gtrsim 100$ mJy of correlated VLBA flux density at 15 GHz, and are derived from several flux-density limited and representative samples.  The MOJAVE 1.5 Jy sample encompasses all 181 AGN above J2000 declination $-30\arcdeg$ known to have exceeded 1.5 Jy in VLBA 15 GHz flux density at any epoch between 1994.0 and 2010.0 \citep{MOJAVE_X}. With the launch of the {\it Fermi} observatory in 2008, two new $\gamma$-ray-based AGN samples were added to MOJAVE.  The 1FM sample \citep{2011ApJ...742...27L} consists of all 116 AGN in the 1FGL catalog \citep{1FGL} above declination $-30\arcdeg$ and galactic latitude $|b| > 10\arcdeg$ with average integrated {\it Fermi} LAT $>0.1$ GeV energy flux above $3 \times 10^{-11}\, \mathrm {erg\,   cm^{-2}\,s^{-1}}$.  In 2013, a new hard spectrum MOJAVE sample of 132 AGN was added, which had declination $> -20\arcdeg$, total 15 GHz VLBA flux density $> 0.1$ Jy, and mean {\it Fermi} 2LAC catalog \citep{2LAC} or 3LAC catalog \citep{2015ApJ...810...14A} spectral index harder than 2.1.

Of the 387 AGN presented in this paper, 304 are members of one or more of the above samples. An additional 78 AGN are members of either the pre-cursor survey to MOJAVE (the 2cm VLBA survey; \citealt{1998AJ....115.1295K}), the low-luminosity MOJAVE AGN sample \citep{MOJAVE_X}, the 3rd EGRET $\gamma$-ray catalog \citep{Hartman99}, the 3FGL {\it Fermi} LAT $\gamma$-ray catalog \citep{3FGL}, or the ROBOPOL optical polarization monitoring sample \citep{2014MNRAS.442.1693P}. Finally, we have included five AGN that were originally candidates for the above samples, but did not meet the final selection criteria. All AGN have 15 GHz VLBI flux density greater than 100 mJy and at least five epochs of observation with polarization.

We gathered observer frame synchrotron peak frequency values from the literature (mainly \citealt{2015AA...578A..69H}) where possible. For roughly half the sample, we used the ASDC SED Builder tool \citep{2011arXiv1103.0749S} to determine the peak frequency from a parabolic fit to published multiwavelength data in the (log $\nu$, log$(\nu F_\nu)$) plane, where $F_\nu$ is the flux density at frequency $\nu$. In the case of 24 sources, there were insufficient available data to determine a synchrotron peak frequency. We assigned a SED peak classification based on the rest frame peak values $\nu_p$ as follows: low-synchrotron-peaked (LSP ; $\nu_p$ below $10^{14}$ Hz); intermediate-synchrotron-peaked (ISP; $\nu_p$  between $10^{14}$ Hz and $10^{15}$ Hz); and high-synchrotron-peaked (HSP; $\nu_p$  above $10^{15}$ Hz). For 46 BL Lacs without known redshifts, and eight optically unidentified sources, we assumed a redshift of 0.3 in calculating the rest frame SED peak value. The known AGN redshifts and their literature references are listed in Table~\ref{gentable}, along with optical class and SED peak classification. Table~\ref{samplesummary} contains a breakdown of sample AGN by optical and SED peak classification. The single epoch polarization results published in \citealt{MOJAVE_I} were based on a flux-density limited sample of 135 AGN, of which four were ISP and the remainder were LSP. This paper covers an additional 252 AGN, for a final total of 312 LSP, 30 ISP, and 21 HSP AGN, with the latter two categories consisting entirely of BL Lacs.

\begin{deluxetable*}{clccccccl} 
\tablecolumns{9} 
\tabletypesize{\scriptsize} 
\tablewidth{0pt}  
\tablecaption{\label{gentable} General Properties of the AGN Sample}  
\tablehead{ & & & & & & \colhead{$\mathrm{log(\nu_{peak})}$} & & \\ 
\colhead{Source} & \colhead{Alias} & \colhead{1.5 Jy} & \colhead{LAT} & \colhead{Optical Class} & \colhead{SED Peak Class} & \colhead{(Hz)} & \colhead{Redshift} & \colhead{Ref.} \\ 
\colhead{(1)} & \colhead{(2)} & \colhead{(3)} & \colhead{(4)} & \colhead{(5)} & \colhead{(6)} & \colhead{(7)} & \colhead{(8)} & \colhead{(9)}} 
\startdata 
0003+380 & S4 0003+38 & N & Y & Q & LSP & 13.2 & 0.229 & \cite{1994AAS..103..349S} \\ 
0003$-$066 & NRAO 005 & Y & N & B & LSP & 13.1 & 0.3467 & \cite{2005PASA...22..277J} \\ 
0006+061 & CRATES J000903+062820 & N & Y & B & LSP & 13.8 & 1.56263 & \cite{2017ApJS..233...25A} \\ 
0007+106 & III Zw 2 & Y & N & G & LSP & 13.3 & 0.0893 & \cite{1970ApJ...160..405S} \\ 
0010+405 & 4C +40.01 & N & N & Q & LSP & 13.0 & 0.256 & \cite{1992ApJS...81....1T} \\ 
0011+189 & RGB J0013+191 & N & Y & B & LSP & 13.9 & 0.477 & \cite{2013ApJ...764..135S} \\ 
0015$-$054 & PMN J0017$-$0512 & N & Y & Q & LSP & 13.7 & 0.226 & \cite{2012ApJ...748...49S} \\ 
0016+731 & S5 0016+73 & Y & N & Q & LSP & 12.7 & 1.781 & \cite{1986AJ.....91..494L} \\ 
0019+058 & PKS 0019+058 & N & Y & B & LSP & 13.2 & \nodata & \cite{2013ApJ...764..135S} \\ 
0027+056 & PKS 0027+056 & N & N & Q & LSP & 12.8 & 1.317 & \cite{1999AJ....117...40S} \\ 
0044+566 & GB6 J0047+5657 & N & Y\tablenotemark{1} & B & \nodata & \nodata & 0.747 & \cite{2005ApJ...626...95S} \\ 
0048$-$071 & OB $-$082 & N & Y & Q & LSP & 13.3 & 1.975 & \cite{1983MNRAS.205..793W} \\ 
0048$-$097 & PKS 0048$-$09 & Y & Y & B & ISP & 14.5 & 0.635 & \cite{2012AA...543A.116L} \\ 
0055+300 & NGC 315 & N & N & G & ISP & 14.0 & 0.0165 & \cite{1999ApJS..121..287H} \\ 
0059+581 & TXS 0059+581 & Y & Y\tablenotemark{1} & Q & LSP & 12.9 & 0.644 & \cite{2005ApJ...626...95S} \\ 
0106+013 & 4C +01.02 & Y & Y & Q & LSP & 13.0 & 2.099 & \cite{1995AJ....109.1498H} \\ 
0106+612 & TXS 0106+612 & N & Y\tablenotemark{1} & Q & LSP & 13.5 & 0.783 & \cite{2010ApJ...718L.166V} \\ 
0106+678 & 4C +67.04 & N & Y\tablenotemark{1} & B & LSP & 15.0 & 0.29 & \cite{2010ApJ...712...14M} \\ 
0109+224 & S2 0109+22 & Y & Y & B & LSP & 13.5 & \nodata & \cite{2017ApJ...837..144P} \\ 
0109+351 & B2 0109+35 & Y & N & Q & LSP & 13.0 & 0.45 & \cite{1996MNRAS.282.1274H} \\ 
0110+318 & 4C +31.03 & N & Y & Q & LSP & 13.1 & 0.603 & \cite{1976ApJS...31..143W} \\ 
0111+021 & UGC 00773 & N & N & B & LSP & 12.7 & 0.047 & \cite{1976ApJS...31..143W} \\ 
0113$-$118 & PKS 0113$-$118 & N & Y & Q & LSP & 13.1 & 0.671 & \cite{2012ApJ...748...49S} \\ 
0116$-$219 & OC $-$228 & N & Y & Q & LSP & 13.2 & 1.165 & \cite{1983MNRAS.205..793W} \\ 
\enddata 
\tablecomments{Columns are as follows: (1) B1950 name, (2) other name, (3) MOJAVE 1.5 Jy catalog inclusion, (4) status of $Fermi$-LAT detection, (5) optical classification where Q = FSRQ, B = BL Lac, G = radio galaxy, N = narrow line Seyfert 1, and U = unidentified, (6) class based on synchrotron peak frequency, (7) rest frame log synchrotron peak frequency in Hz (for sources with no known redshift, z=0.3 was used), (8) redshift, (9) reference for redshift (or optical class, for BL Lacs without known redshift).}
\tablenotetext{1}{Located within 10 degrees of the galactic plane; omitted from $\gamma$-ray comparisons.}
(Only a portion of this table is shown here to demonstrate its form and content. A machine-readable version of the full table is available.)
\end{deluxetable*}

\begin{deluxetable}{lcccccc}
\tablecolumns{6} 
\tabletypesize{\scriptsize} 
\tablewidth{0pt}  
\tablecaption{\label{samplesummary} Optical and Synchrotron Peak Classifications}  
\tablehead{\colhead{Optical Class} & \colhead{LSP} & \colhead{ISP} & \colhead{HSP} & \colhead{Unknown} & \colhead{Total}} 
	\startdata 
	FSRQ & 223 & 0 & 0 & 16 & 239 \\ 
	BL Lac & 64 & 30 & 21 & 3 & 118 \\ 
	Radio Galaxy & 14 & 0 & 0 & 3 & 17 \\
	NLSy1 &  5 & 0 & 0 & 0 & 5 \\ 
	Unidentified & 6 & 0 & 0 & 2 & 8 \\ 
    Total & 312 & 30 & 21 & 24 & 387 \\
	\enddata 
\end{deluxetable} 

We reduced the polarimetric data (listed in Table~\ref{meastable}) according to the procedures described in \citealt{MOJAVE_I} and \citealt{2018ApJS..234...12L}. The core was in almost all cases distinguishable as the most compact feature in the radio map, located at the bright end of the jet (see the Appendix and \citealt{MOJAVE_XIII} for notes on individual sources). We determined its position by fitting Gaussians to the \((u,v)\) visibility data using Difmap \citep{difmap}. The mean $I$, $Q$, and $U$ of the nine contiguous 0.1 mas wide image pixels centered on the core were then used to determine the core's linear fractional polarization and EVPA. The fractional polarization was considered an upper limit equal to five times the $P$ rms noise level divided by $I$ when $P$ fell below the lowest contour of the relevant image presented in \citealt{2018ApJS..234...12L}. The latter were chosen at a level where the amount of spurious blank sky polarization was minimal. Approximately 12\% of the available fractional polarization measurements are upper limits, and thus have an unknown corresponding EVPA. The uncertainty in fractional polarization is approximately 7\% of the given values, due to uncertainties of 5\% in I and P, and the EVPA is accurate within $5\arcdeg$, based on comparisons with near-simultaneous single-dish measurements taken at The University of Michigan Radio Astronomy Observatory (UMRAO; \citealt{2003ApJ...586...33A}). A previous MOJAVE study found that the core rotation measures are variable in time, and at 15 GHz, the median core EVPA rotation was less than 4$\arcdeg$ \citep{2012AJ....144..105H}. For these reasons, we do not correct our measurements for Faraday rotation. Detailed description of the observational sampling is available in \citealt{2018ApJS..234...12L}.

\begin{deluxetable}{cccrrr} 
\tablecolumns{6} 
\tabletypesize{\scriptsize} 
\tablewidth{0pt}  
\tablecaption{\label{meastable} VLBA 15 GHz Core Measurements}  
\tablehead{ & & \colhead{\textit{I}} & \colhead{\textit{m}} & \colhead{EVPA} & \colhead{PA} \\ 
\colhead{Source} & \colhead{Epoch} & \colhead{(Jy beam$^{-1}$)} & \colhead{(\%)} & \colhead{(deg)} & \colhead{(deg)} \\ 
\colhead{(1)} & \colhead{(2)} & \colhead{(3)} & \colhead{(4)} & \colhead{(5)} & \colhead{(6)}} 
\startdata 
0003+380 & 2006-03-09 & 0.52 & 2.5 & 81 & 110 \\ 
  & 2006-12-01 & 0.32 & 0.8 & 130 & 112 \\ 
  & 2007-03-28 & 0.39 & $<$0.3 & \nodata & 115 \\ 
  & 2007-08-24 & 0.39 & $<$0.2 & \nodata & 117 \\ 
  & 2008-05-01 & 0.55 & 0.7 & 180 & 118 \\ 
  & 2008-07-17 & 0.50 & 0.9 & 182 & 119 \\ 
  & 2009-03-25 & 0.31 & 0.9 & 138 & 118 \\ 
  & 2010-07-12 & 0.31 & 1.4 & 137 & 120 \\ 
  & 2011-06-06 & 0.45 & 0.3 & 104 & 115 \\ 
  & 2013-08-12 & 0.53 & 1.1 & 135 & 116 \\ 
0003$-$066 & 2003-02-05 & 1.32 & 5.1 & 14 & 351 \\ 
  & 2004-06-11 & 1.16 & 7.8 & 20 & 10 \\ 
  & 2005-03-23 & 0.87 & 4.9 & 8 & 5 \\ 
  & 2005-06-03 & 0.84 & 4.1 & 11 & 4 \\ 
  & 2005-09-16 & 0.76 & 4.5 & 3 & 2 \\ 
  & 2006-07-07 & 0.65 & 2.2 & $-$4 & 358 \\ 
  & 2007-01-06 & 0.66 & 2.9 & 2 & 356 \\ 
  & 2007-04-18 & 0.73 & 3.7 & 16 & 358 \\ 
  & 2007-07-03 & 0.65 & 2.7 & 17 & 358 \\ 
  & 2007-09-06 & 0.65 & 2.9 & 21 & 359 \\ 
  & 2008-07-30 & 0.88 & 1.9 & 2 & 353 \\ 
  & 2009-05-02 & 1.11 & 6.5 & 5 & 355 \\ 
  & 2009-10-27 & 1.04 & 8.1 & 21 & 0 \\ 
  & 2010-08-06 & 0.87 & 9.0 & 22 & 2 \\ 
\enddata 
\tablecomments{Columns are as follows: (1) B1950 name, (2) observation date, (3) Stokes I intensity in Jy/beam, (4) fractional polarization in \%, (5) electric vector position angle in degrees, (6) inner jet position angle in degrees; unless otherwise noted, measured as the position angle of the closest downstream Gaussian jet component with respect to the core.}
\tablenotetext{\mathrm{a}}{PA measurement method: Flux-weighted position angle average of clean components near the core.}
(Only a portion of this table is shown here to demonstrate its form and content. A machine-readable version of the full table is available.)
\end{deluxetable}

\section{Data Analysis and Discussion}

\subsection{Statistical Tests} \label{stattests}

The MOJAVE program observes each individual AGN with a cadence that is appropriate for its angular jet expansion speed, thus some AGN are more densely sampled in time than others. The unequal span of coverage time and frequency of sampling therefore affects some statistics; for example, any measures of maximum polarization or of flux variability are dependent on how long and how frequently the AGN has been observed, since greater time coverage creates more opportunity for the AGN to be observed in a flaring state. On the other hand, median statistics ($m_\mathrm{med}$, $I_\mathrm{med}$, and $|EVPA-PA|_\mathrm{med}$) are less dependent on these factors and so we calculate them using all available epochs on the source. For all other derived quantities, we compensate for sampling biases by using only five epochs of observation spanning a standardized time length of 2.3 years. This time range represents the median total coverage time of all the AGN which had only five polarization image epochs. For each source we use the most recent period with at least five epochs of observation; when a source has more than five observations during that time range, we choose five epochs at random and use these consistently for each calculation. Of the 387 AGN in the sample, 86 do not have a period which qualifies, i.e., at least five observations within any 2.3 year period. We further omit AGN from statistical quantities where five epoch measurements are not available to be used in the calculation - for example, the EVPA statistics require five epochs with measured polarization values (not upper limits). We tabulate all the derived quantities in Table~\ref{derivedtable}. We computed two-tailed Kolgoromov-Smirnoff (KS) tests and correlation tests with Kendall's tau coefficient using the R core package \citep{Rcore}; for censored data comparisons we used the Peto \& Peto modification of the Gehan-Wilcoxon test \citep{RNADA}. In what follows, "significant" statistical test results refer to those whose chance probability is below 5\%.

\begin{deluxetable*}{ccccccccc} 
\tablecolumns{9} 
\tabletypesize{\scriptsize} 
\tablewidth{0pt}  
\tablecaption{\label{derivedtable} Derived Core Quantities}  
\tablehead{ & \colhead{$m_\mathrm{med}$} & \colhead{$m_\mathrm{max}$} & \colhead{$\mathrm{\textit{m}_{var}}$} & \colhead{$\mathrm{\textit{I}_{med}}$} & \colhead{$\mathrm{\textit{I}_{var}}$} & \colhead{$EVPA_\mathrm{{var}}$} & \colhead{$(EVPA-PA)_\mathrm{{var}}$} & \colhead{$|EVPA-PA|_\mathrm{med}$} \\
\colhead{Source} & \colhead{(\%)} & \colhead{(\%)} & & \colhead{(Jy beam$^{-1}$)} & & \colhead{(deg)} & \colhead{(deg)} & \colhead{(deg)} \\ 
\colhead{(1)} & \colhead{(2)} & \colhead{(3)} & \colhead{(4)} & \colhead{(5)} & \colhead{(6)} & \colhead{(7)} & \colhead{(8)} & \colhead{(9)}} 
\startdata
0003+380 & 0.8 & 0.9 & \nodata & 0.42 & 0.22 & \nodata & \nodata & 19 \\ 
0003$-$066 & 5.0 & 9.0 & 0.06 & 0.75 & 0.08 & 8 & 10 & 10 \\ 
0006+061 & 2.1 & 4.5 & 0.49 & 0.13 & 0.10 & 5 & 4 & 17 \\ 
0007+106 & \nodata & \nodata & \nodata & 0.67 & \nodata & \nodata & \nodata & 66 \\ 
0010+405 & \nodata & 0.3 & \nodata & 0.47 & 0.11 & \nodata & \nodata & 24 \\ 
0011+189 & 3.6 & 4.6 & 0.09 & 0.10 & 0.05 & 10 & 9 & 15 \\ 
0015$-$054 & 1.6 & 2.1 & 0.21 & 0.20 & 0.43 & 44 & 38 & 43 \\ 
0016+731 & 1.8 & \nodata & \nodata & 1.21 & \nodata & \nodata & \nodata & 30 \\ 
0019+058 & 0.8 & 3.0 & 0.90 & 0.25 & 0.08 & 30 & 32 & 48 \\ 
0027+056 & 2.1 & 2.7 & 0.30 & 0.44 & 0.10 & 8 & 8 & 53 \\ 
0044+566 & 6.5 & 13.9 & 0.42 & 0.14 & 0.10 & 4 & 4 & 70 \\ 
0048$-$071 & 1.9 & 3.1 & 0.34 & 0.84 & 0.01 & 8 & 8 & 19 \\ 
0048$-$097 & 1.8 & 2.8 & \nodata & 0.64 & 0.63 & \nodata & \nodata & 35 \\ 
0055+300 & \nodata & \nodata & \nodata & 0.20 & \nodata & \nodata & \nodata & \nodata \\ 
0059+581 & 1.7 & 3.2 & 0.43 & 2.20 & 0.18 & 26 & 32 & 26 \\ 
0106+013 & 1.8 & \nodata & \nodata & 1.44 & \nodata & \nodata & \nodata & 56 \\ 
0106+612 & 1.0 & 3.2 & 0.70 & 0.34 & 0.27 & 24 & 25 & 70 \\ 
0106+678 & 2.0 & \nodata & \nodata & 0.15 & \nodata & \nodata & \nodata & 8 \\ 
0109+224 & 3.1 & 3.2 & 0.78 & 0.43 & 0.20 & 43 & 46 & 17 \\ 
0109+351 & 1.1 & \nodata & \nodata & 0.57 & \nodata & \nodata & \nodata & 76 \\ 
0110+318 & 1.0 & 3.5 & 0.69 & 0.52 & 0.06 & 48 & 48 & 33 \\ 
0111+021 & \nodata & \nodata & \nodata & 0.29 & \nodata & \nodata & \nodata & \nodata \\ 
0113$-$118 & 0.4 & \nodata & \nodata & 0.74 & \nodata & \nodata & \nodata & 10 \\ 
0116$-$219 & 1.1 & 2.5 & 0.37 & 0.46 & 0.30 & 26 & 26 & 36 \\ 
0118$-$272 & 5.0 & \nodata & \nodata & 0.18 & \nodata & \nodata & \nodata & 2 \\ 
0119+115 & 3.7 & \nodata & \nodata & 0.97 & \nodata & \nodata & \nodata & 9 \\ 
0122$-$003 & 0.8 & 1.0 & \nodata & 0.29 & 0.15 & \nodata & \nodata & 72 \\ 
0130$-$171 & 1.9 & 2.5 & 0.74 & 1.46 & 0.32 & 22 & 20 & 70 \\ 
\enddata 
\tablecomments{Columns are as follows: (1) B1950 name, (2) median fractional polarization in \%, (3) maximum fractional polarization in \%, (4) fractional polarization variability, (5) median total (Stokes $I$) intensity in Jy/beam, (6) total intensity variability, (7) EVPA variability in degrees, (8) EVPA variability computed relative to jet PA in degrees, (9) median absolute difference between EVPA and jet PA in degrees.}
(Only a portion of this table is shown here to demonstrate its form and content. A machine-readable version of the full table is available.)
\end{deluxetable*}

\subsection{Fractional Polarization} \label{fracpol}

\begin{figure}
	\plotone{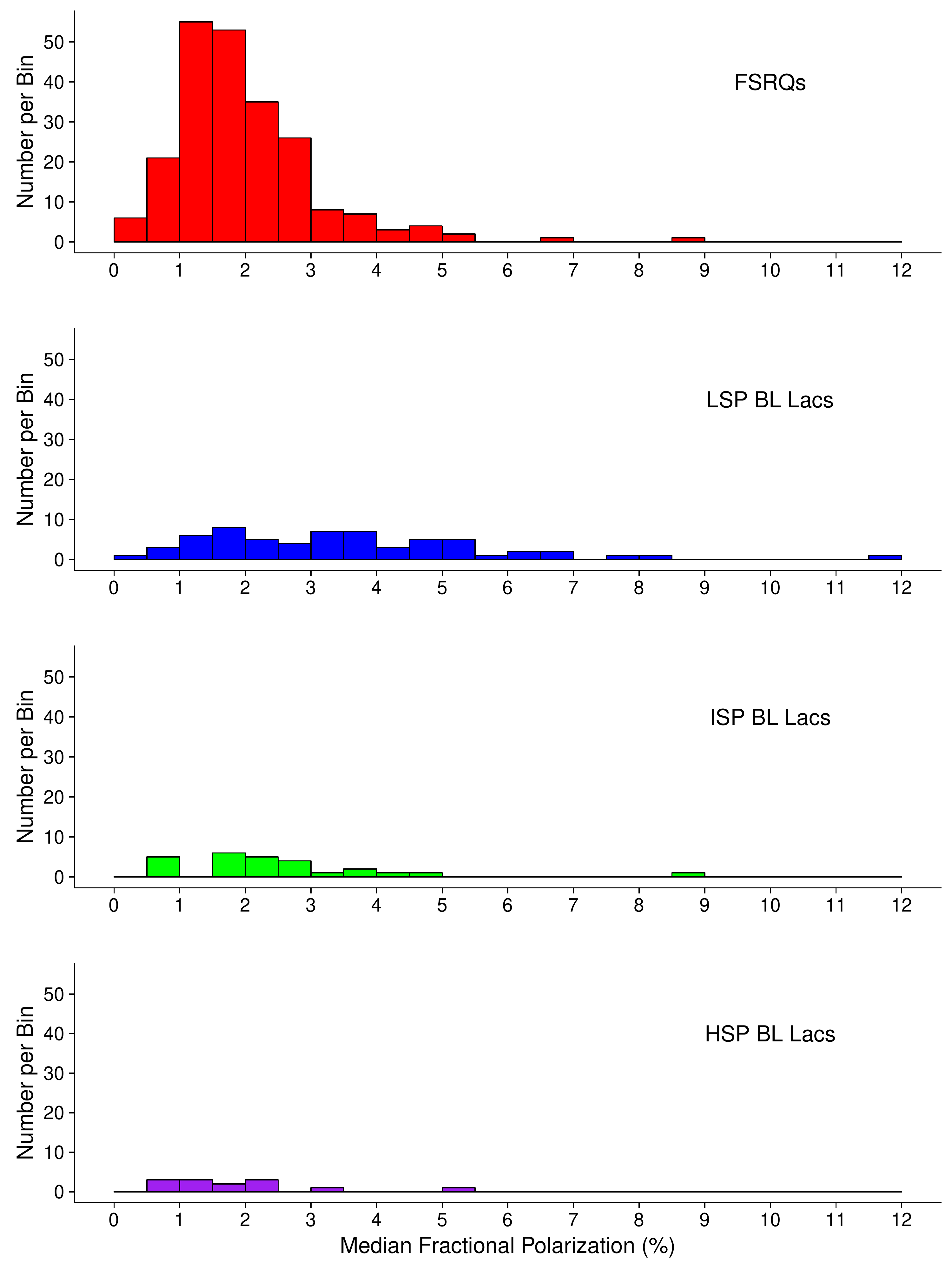}
    \caption{\label{medms} Distributions of median fractional polarization $m_\mathrm{med}$, grouped by optical/synchrotron peak classification.}
\end{figure}

\begin{figure}
	\plotone{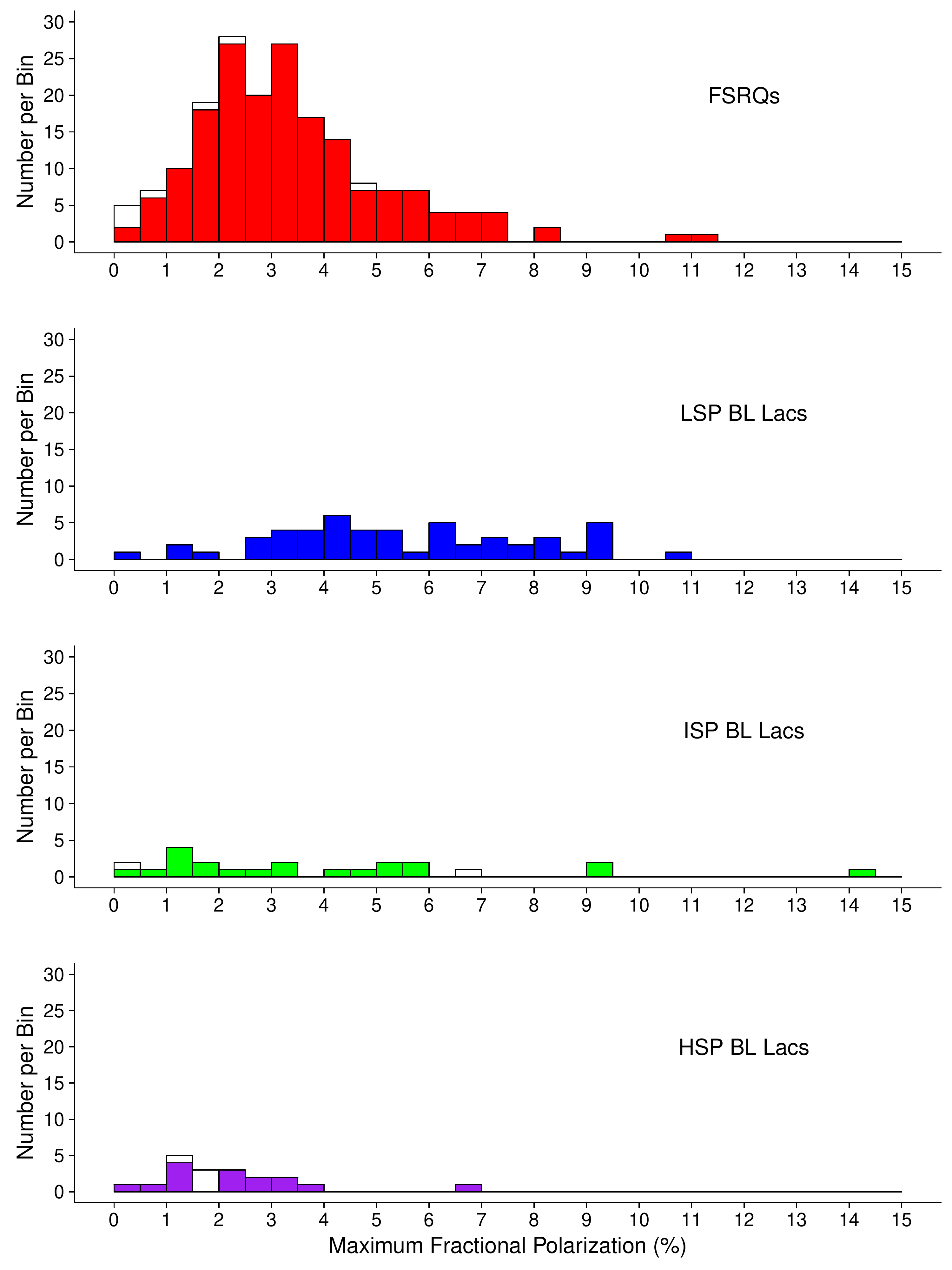}
    \caption{\label{maxms} Distributions of maximum fractional polarization $m_\mathrm{max}$, grouped by optical/synchrotron peak classification. Unfilled bins represent upper limit measurements.}
\end{figure}

We describe the magnitude of polarization of each AGN core with two statistical quantities: the median fractional polarization ($m_\mathrm{med}$) and the maximum fractional polarization ($m_\mathrm{max}$). In Figure~\ref{medms}, we show distributions of median fractional polarization, over all epochs of each AGN, for the most common categories of optical/SED class (50 AGN are omitted due to too many upper limit measurements to ascertain a confident median; an additional five narrow line Seyfert 1 galaxies and nine AGN of unknown optical class are not shown). The maximum fractional polarization distributions of the BL Lacs and FSRQs are shown in Figure~\ref{maxms}.

\begin{figure}
	\plotone{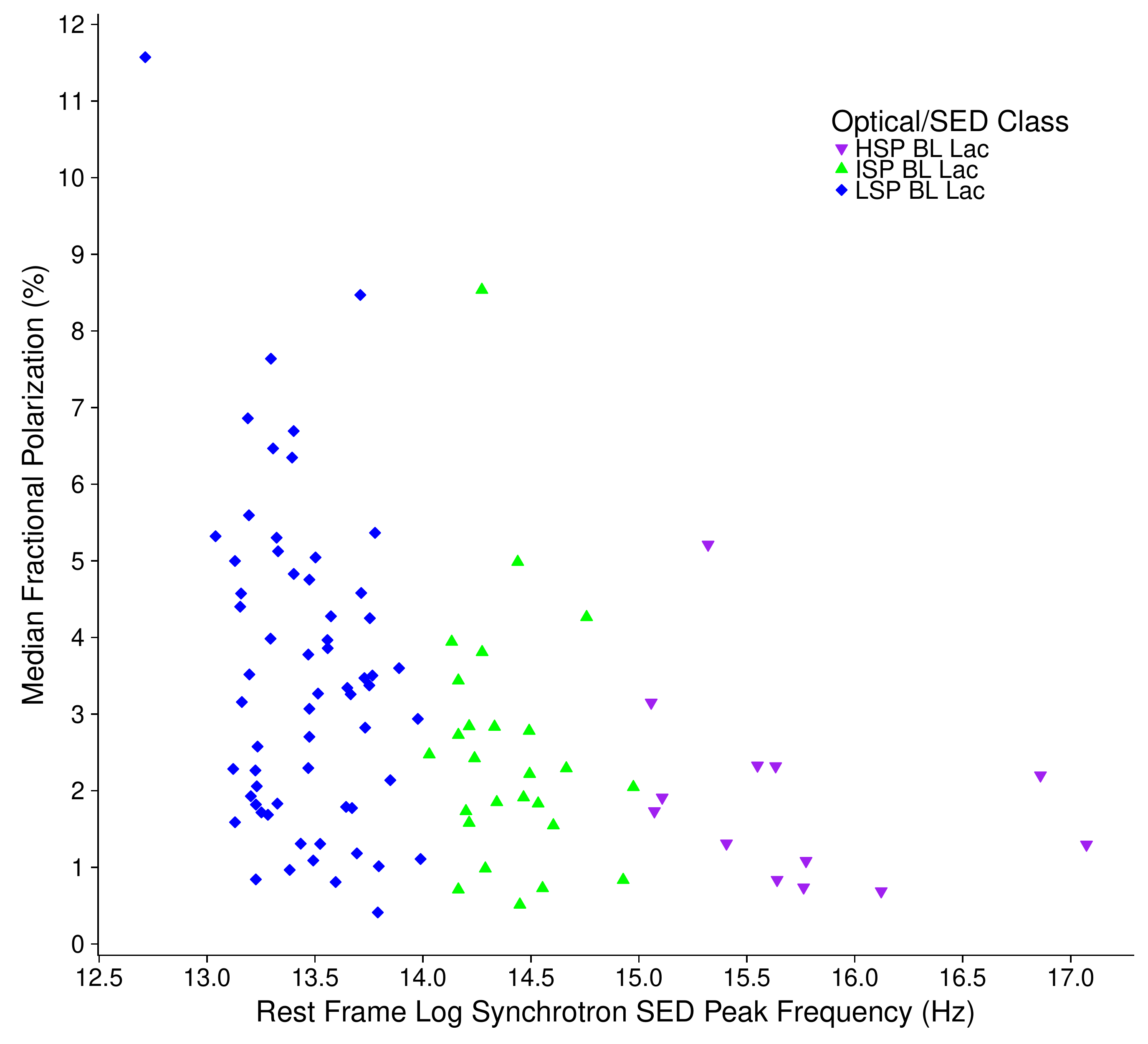}
	\caption{\label{bllseds} Rest frame log synchrotron SED peak frequency versus median fractional polarization $m_\mathrm{med}$ for BL Lacs in the sample. Purple inverted triangles are HSP BL Lacs, green triangles are ISP BL Lacs, and blue diamonds are LSP BL Lacs. From left to right, the most polarized outliers in each class are S5 0346+80, GB6 J0929+5013, and GB6 J0154+0823. A Kendall tau test of correlation yields $p=9.0\times 10^{-5}$ for no correlation.}
\end{figure}

\begin{figure}
	\plotone{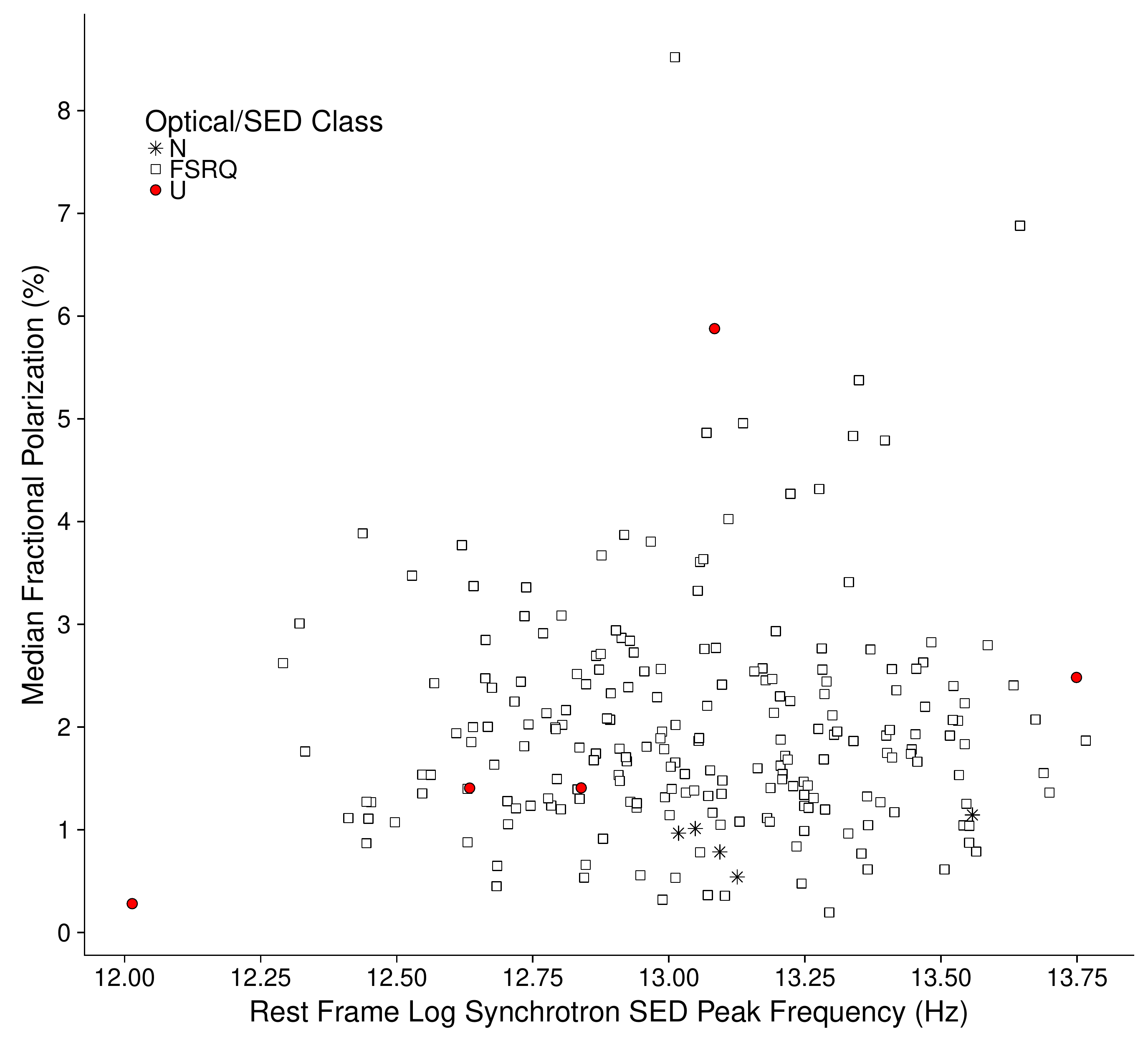}
	\caption{\label{otherseds} Rest frame log synchrotron SED peak frequency versus median fractional polarization $m_\mathrm{med}$ for non-BL Lacs in the sample. Black stars are NLSy1s, unfilled squares are FSRQs, and red circles have an unknown optical class. The highly polarized outlier is PKS 1236+077. There is no significant correlation.}
\end{figure}

LSP BL Lacs are the most fractionally polarized, with an average median over 3\%, and are significantly different from both HSP BL Lacs (KS test result $p=0.0097$) and FSRQs (KS test result $p=2.7\times 10^{-9}$). HSP BL Lacs and FSRQs are similar, however, and each of their median fractional polarization distributions has a mean below $\sim$2\%. The median fractional polarization of BL Lacs is strongly anti-correlated with synchrotron SED peak frequency (shown in Figure~\ref{bllseds}; other optical classes are shown in Figure~\ref{otherseds}). This trend for BL Lacs was previously reported by \citealt{2011ApJ...742...27L} and \citealt{2012ApJ...757...25L}, of which the former theorized that AGN with lower synchrotron peaks may be more Doppler boosted than their high synchrotron peaked counterparts, resulting in higher core polarization. This was partially based on the low core brightness temperatures of HSP BL Lacs, a relationship which still holds true (Homan, D. C. et al., in preparation). An alternate explanation proposed by \citealt{2016MNRAS.463.3365A} is based on a shock-in-jet model. In this scenario, the volume immediately downstream of the shock is the origin of emission near the synchrotron SED peak frequency and above. If the shock orders an underlying magnetic field, then higher polarization might be expected closer to these frequencies. While \citealt{2016MNRAS.463.3365A} use this theory to describe the same anti-correlation in optical polarization, at 15 GHz we are also probing closer to the peak frequency of LSP BL Lacs than HSP BL Lacs, which could explain the lower polarization of the latter.

In \citealt{2011ApJ...742...27L} beam depolarization effects were given as a possible reason for the difference between LSP BL Lac and FSRQs, as the former have much lower redshifts on average. In our sample, there is a cohort of seven high-redshift FSRQs with fractional polarization which is lower than average, but no trend between redshift and fractional polarization exists below $z=2.5$. The highest redshift FSRQs alone cannot therefore be responsible for the relatively low polarization of the class. In \citealt{MOJAVE_I} and \citealt{ALL99}, the (mostly LSP) BL Lacs also had greater core polarization than FSRQs; these authors suggested the cause was slower flow speeds in BL Lacs. Slower flows could create stronger transverse shocks compared to oblique ones in FSRQs, leading to a more ordered magnetic field and better alignment between the EVPA and the local jet direction. We note that $m_\mathrm{med}$ is unavailable for all of the 17 radio galaxies in our sample, of which all but III Zw 2 have at least half their core $m$ measurements classified as upper limits. This suggests that their polarization as a class is low; in contrast, $m_\mathrm{med}$ is available for $\sim$60$\%$ of HSP BL Lacs despite the fact that their $I$ is typically lower than that of the radio galaxies. This low polarization is likely due to higher angles to the line of sight \citep{2013EPJWC..6106001W}.

\subsection{EVPA Alignment With the Inner Jet}

In order to investigate possible alignments between the core EVPA direction and each jet, we determined an inner jet position angle (jet PA) at each epoch using the VLBA data.  For the majority of AGN, we used the method described in \citealt{MOJAVE_X}, which takes a flux-weighted position angle average of the components near the core generated by the CLEAN imaging algorithm \citep{1974A&AS...15..417H}. This method fails in cases where there is an emission gap between the core and downstream features; in the case of all epochs of 39 AGN we therefore used the position angle of the closest downstream Gaussian-fitted jet feature, as was done in \citealt{MOJAVE_I}.  In some individual epoch images, it was impossible to determine a jet PA due to a complete lack of apparent downstream jet emission.  These are indicated in Table~\ref{meastable}.

\begin{figure}
	\plotone{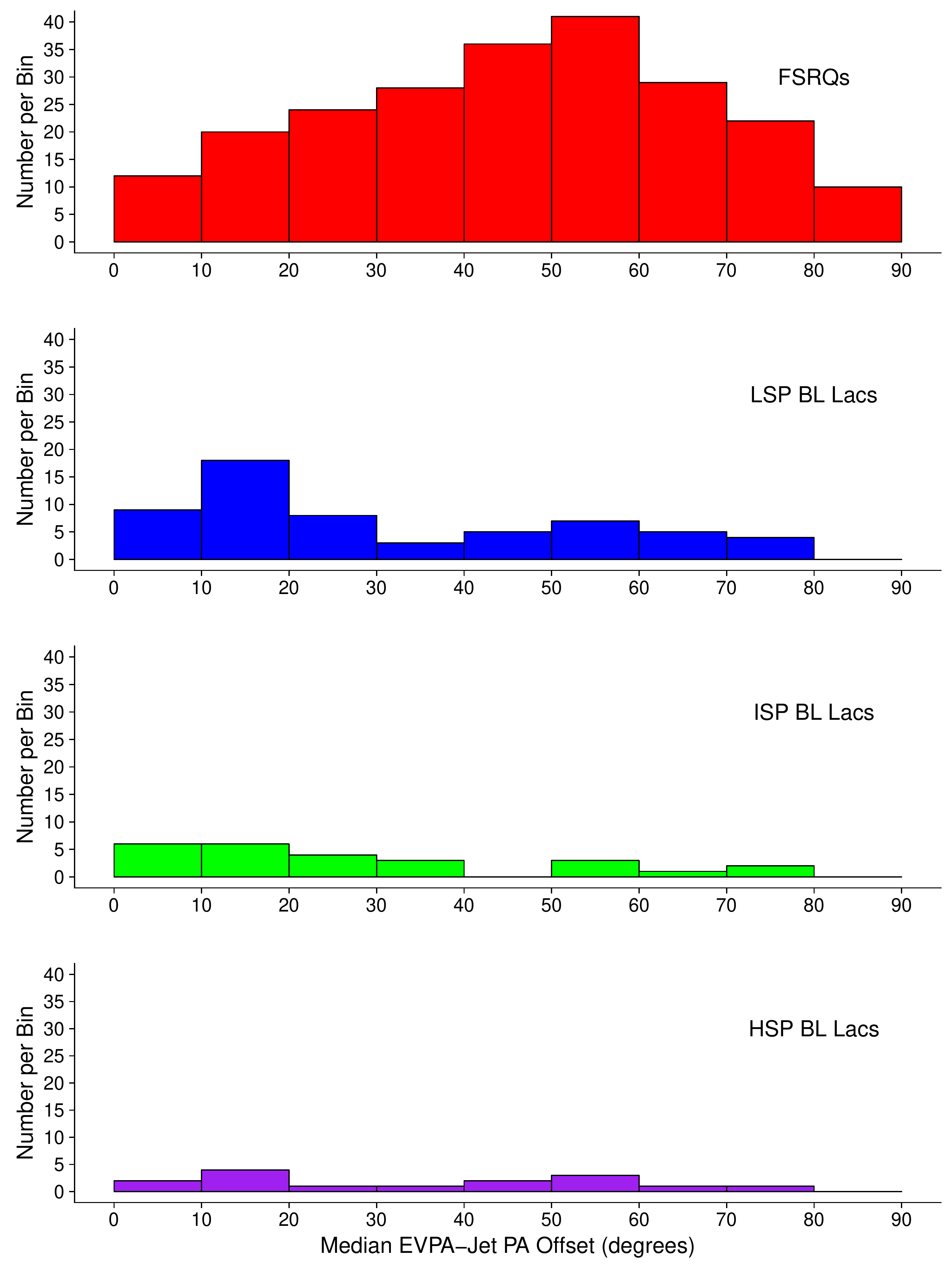}
    \caption{\label{medevpapa} Distributions of median EVPA-jet PA offset $|EVPA-PA|_\mathrm{med}$, grouped by optical/synchrotron peak classification.}
\end{figure}

For the epochs described in Section~\ref{stattests}, we rotated the EVPA to within $90\arcdeg$ of the jet PA (by adding or subtracting $180\arcdeg$), subtracted it from the PA, and took the absolute value of this angle difference to get an EVPA offset value. The distribution of median offsets for each AGN class are shown in Figure~\ref{medevpapa}. The EVPAs of BL Lacs are more often aligned with the local jet direction (i.e., low offset values), in contrast to FSRQs. A KS test between the LSP BL Lac and FSRQ offset distributions returns a p-value of 1.69$\times 10^{-5}$ that they derive from the same parent population.  None of the BL Lac subclass distributions are significantly different from each other. These findings on the difference in alignment between FSRQs and BL Lacs in general agree with the single-epoch results presented in \citealt{MOJAVE_I}, which contained only four ISP and no HSP AGN. Beyond the aforementioned shocks, another suggested scenario is that both types of jets possess a helical magnetic field; the slower flow speeds of BL Lacs could lead to a different pitch angle such that they have a stronger toroidal component to their magnetic field, leading to better EVPA-jet alignment compared to FSRQs \citep{2000MNRAS.319.1109G, 2002PASJ...54L..39A}. It should be mentioned, though, that various other VLBI studies of radio core EVPA/PA alignment in AGN jets have reported contradictory findings on whether alignment, anti-alignment, or both cases are expected (see \citealt{2018MNRAS.473.1850A} and references therein).

\begin{figure}
	\plotone{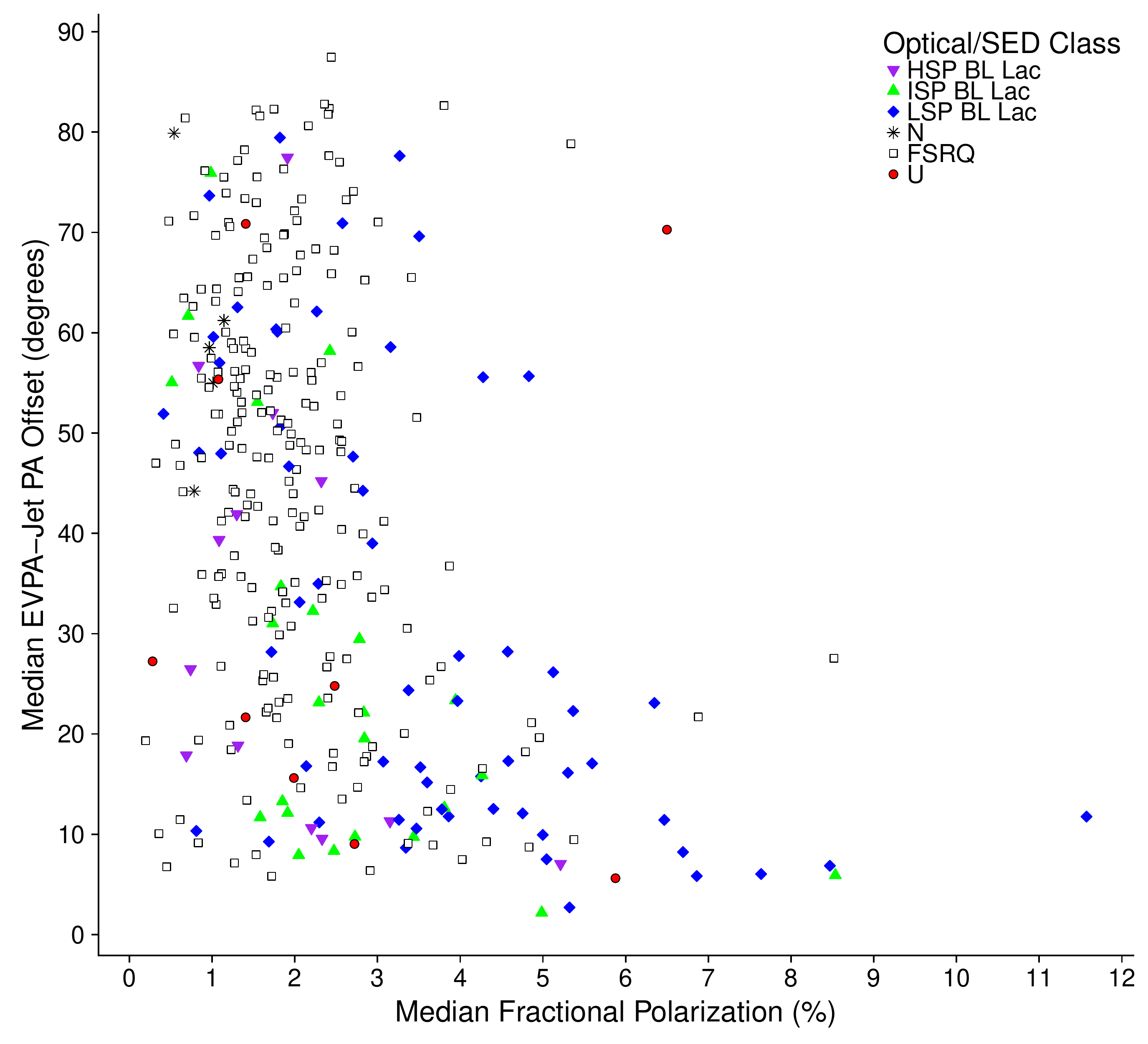}
	\caption{\label{medmmedevpapa} Median fractional polarization $m_\mathrm{med}$ versus median EVPA-jet PA offset $|EVPA-PA|_\mathrm{med}$ for each AGN over time. Purple inverted triangles are HSP BL Lacs, green triangles are ISP BL Lacs, blue diamonds are LSP BL Lacs, black stars are NLSy1s, unfilled squares are FSRQs, and red circles have an unknown optical class or synchrotron peak frequency. A Kendall tau test of correlation yields $p=5.3\times 10^{-13}$ for no correlation.}
\end{figure}

Of the studies suggesting an anti-alignment between core EVPAs and the jet direction (e.g., \citealt{2007ApJ...658..203H}), some  at 5 GHz have found that only FSRQs, not BL Lacs, prefer an EVPA-jet PA offset of near $90\arcdeg$ \citep{2003ApJ...589..733P, 2003ApJ...586...33A}. Our data show no apparent clustering at $90\arcdeg$, but the FSRQ distribution is skewed towards misalignment (median $|EVPA-PA|_\mathrm{med}$=$48\arcdeg$). Figure~\ref{medmmedevpapa} shows that the most fractionally polarized AGN cores, mainly LSP BL Lacs, tend to have EVPAs closely aligned with the local jet direction - a trend which persists if each optical class is inspected individually. This supports a scenario where the magnetic field is more ordered when transverse to the jet. We further discuss the EVPA/PA and their time evolution in Section~\ref{evpavar}.

\subsection{Fractional Polarization Variability}

Because of the stochastic nature of AGN light curves, there has been no clear consensus on what constitutes an ideal statistical measure of flux or fractional polarization variability over time, as discussed by \citealt{2014MNRAS.438.3058R}. Many commonly-used statistics depend on frequent and/or regular observations of all sources, and are therefore not well-suited to the MOJAVE data. We use the variability index of \citealt{ALL99}, and later adopted by \citealt{2007AJ....134..799J}, defined by: \begin{equation} \label{var} m_\mathrm{var} = \frac{(m_\mathrm{max}-\sigma_\mathrm{max})-(m_\mathrm{min}+\sigma_\mathrm{min})}{(m_\mathrm{max}-\sigma_\mathrm{max})+(m_\mathrm{min}+\sigma_\mathrm{min})}. \end{equation} The error for a particular epoch, $\sigma$, is approximately 7$\%$, based on the definition of fractional polarization and the 5$\%$ errors in $P$ and $I$ as described in \citealt{2018ApJS..234...12L}.  As discussed by \citealt{2003ApJ...586...33A}, this measure is not accurate when the minimum values fall below the signal to noise threshold. We therefore omit any AGN where one or more of the five fractional polarization measurements discussed in Section~\ref{stattests} is an upper limit. We also throw out any cases where the errors are high enough to result in a negative variability index.

\begin{figure}
	\plotone{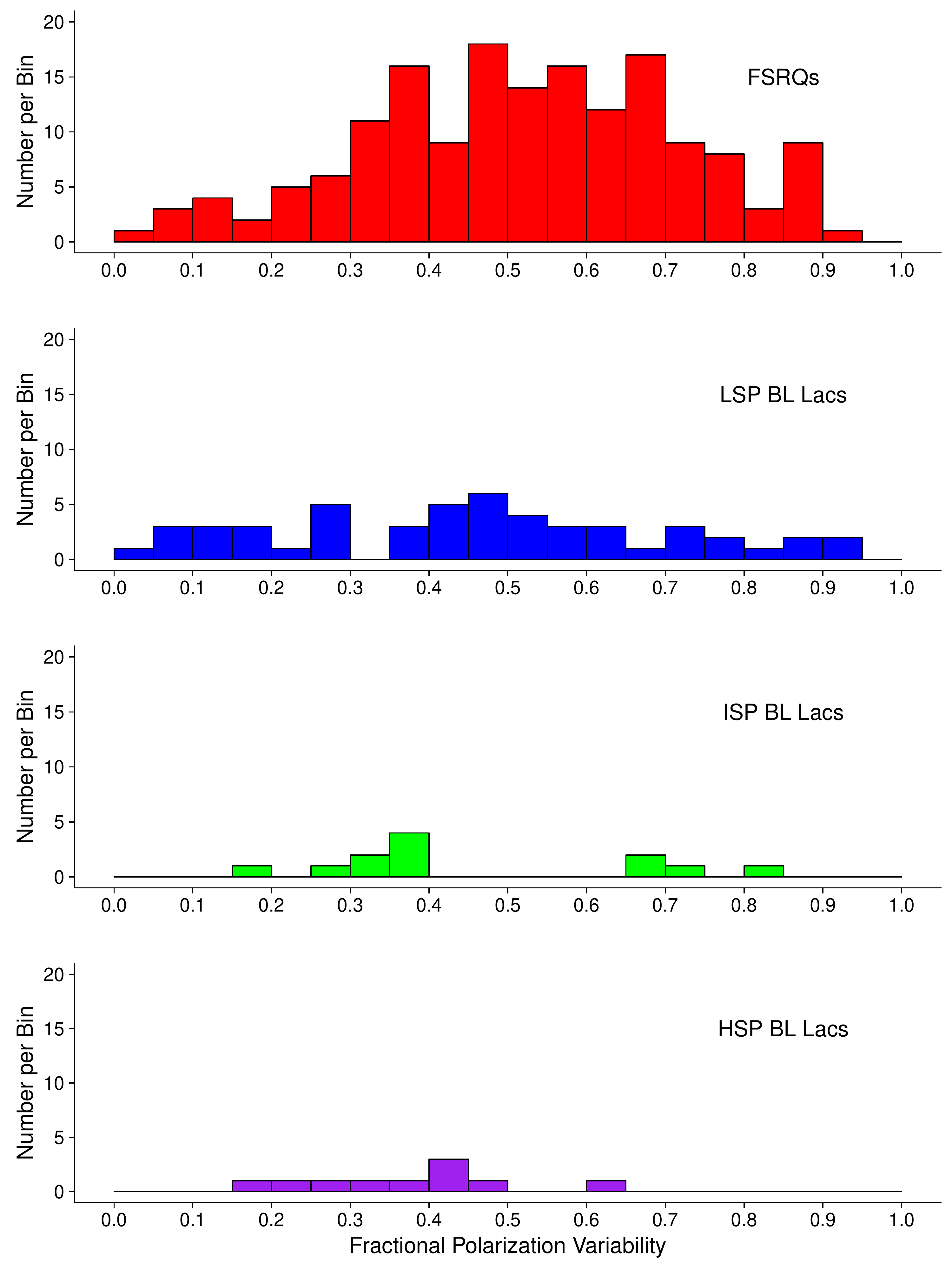}
    \caption{\label{varms} Distributions of fractional polarization variability $m_\mathrm{var}$, grouped by optical/synchrotron peak classification.}
\end{figure}

\begin{figure}
	\plotone{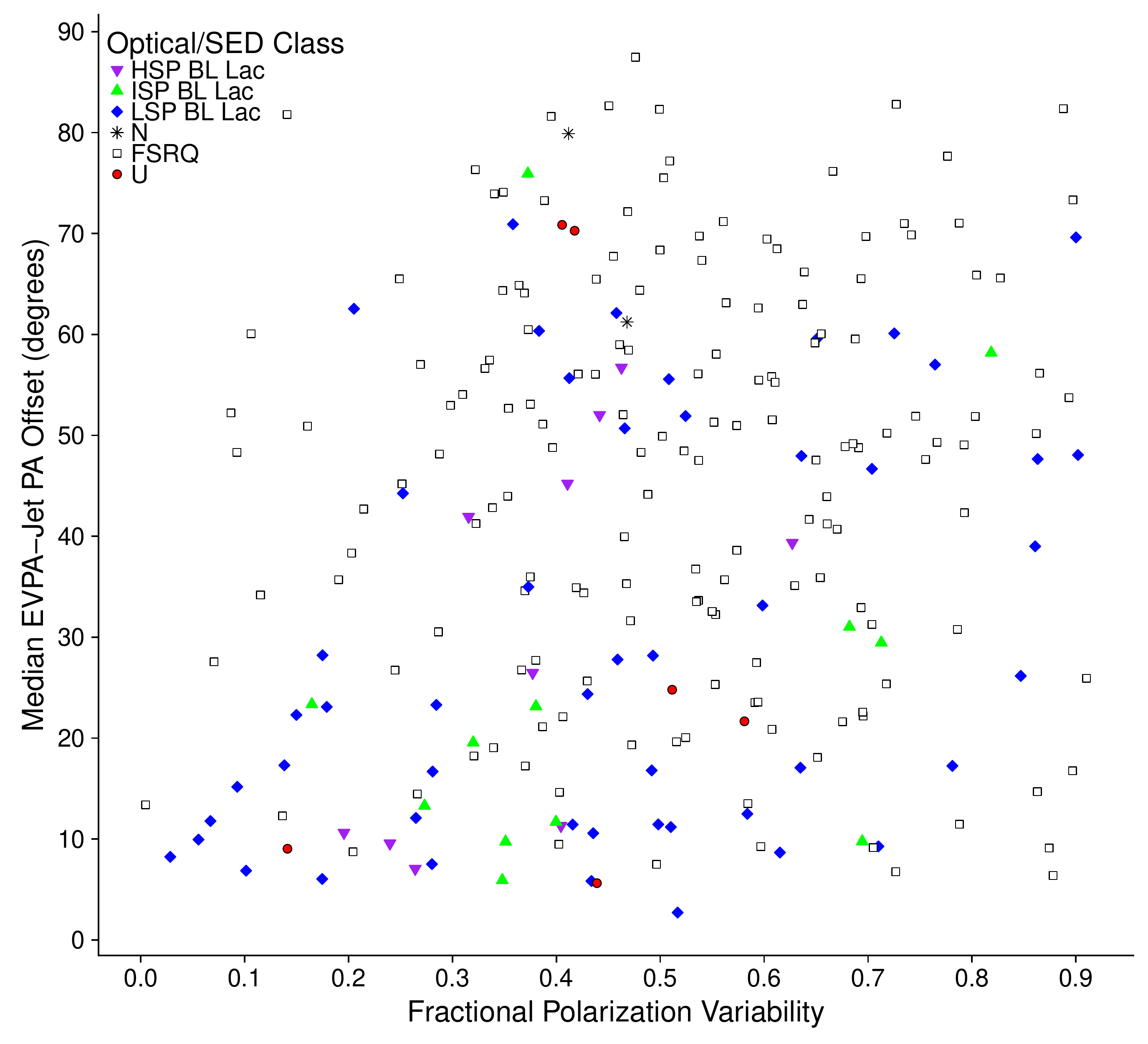}
  	\caption{\label{varmmedabs} Variability in fractional polarization $m_\mathrm{var}$ versus median EVPA-jet PA offset $|EVPA-PA|_\mathrm{med}$ for each AGN over time. Purple inverted triangles are HSP BL Lacs, green triangles are ISP BL Lacs, blue diamonds are LSP BL Lacs, black stars are NLSy1s, unfilled squares are FSRQs, and red circles have an unknown optical class or synchrotron peak frequency. A Kendall tau test of correlation yields $p=0.009$ for no correlation.}
\end{figure}

The variability of fractional polarization (Figure~\ref{varms}) is not significantly dependent on synchrotron peak frequency, although we note that FSRQs appear more variable than HSP BL Lacs (KS test result $p=0.005$). Studies of fractional polarization variability tend to be rare, although \citealt{2007AJ....134..799J} found FSRQs cores to be more variable than (mostly LSP) BL Lacs in a small sample of 15 AGN at 43 GHz. The AGN cores with the smallest fractional polarization variability are more likely to have small EVPA-jet misalignments (Figure~\ref{varmmedabs}).

\subsection{Total Intensity Variability}

\begin{figure}
	\plotone{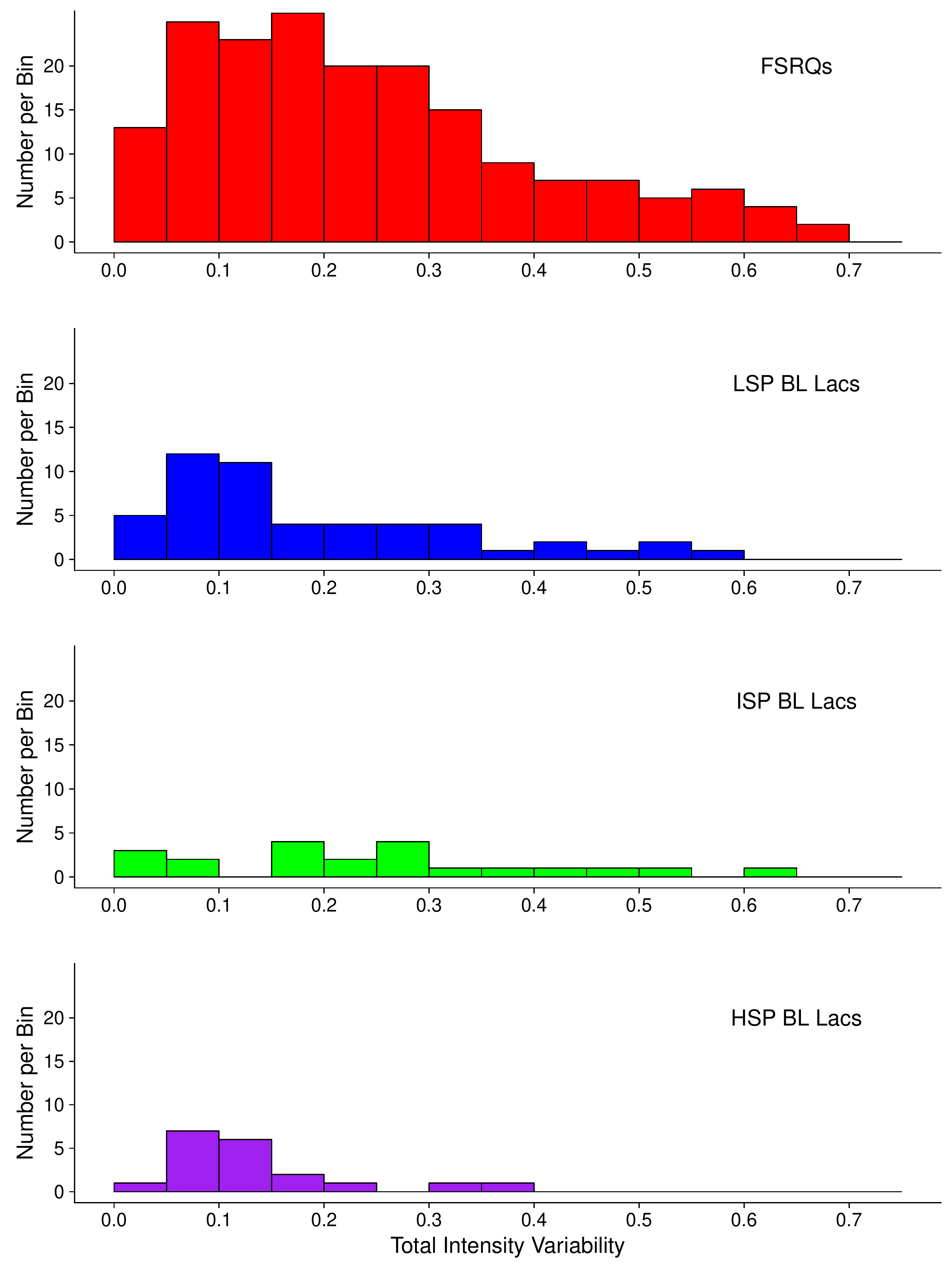}
    \caption{\label{varis} Distributions of total intensity variability $I_{var}$, grouped by optical/synchrotron peak classification.}
\end{figure}

We also use Equation~\ref{var}, replacing $m$ with Stokes $I$, to derive the variability index $I_\mathrm{var}$ for the total intensity. Figure~\ref{varis} shows the $I_\mathrm{var}$ distributions for FSRQs and BL Lac subclasses. Interestingly, while HSP and LSP BL Lacs have similar distributions, ISP BL Lacs peak at a higher variability index---they are significantly different from HSP BL Lacs ($p=0.009$) and nearly so for LSP BL Lacs ($p=0.06$). With the exception of ISP BL Lacs, FSRQs are the most variable blazar class. Tests for a similar parent distribution produced results of $p=0.03$ for FSRQs vs. LSP BL Lacs and $p=0.002$ for FSRQs vs. HSP BL Lacs. Past radio studies of total intensity variability have shown somewhat mixed results; \citealt{2003ApJ...586...33A} also found no significant differences between FSRQs and BL Lacs in their single dish, 5-15 GHz UMRAO study spanning 1984-2001. \citealt{2014MNRAS.438.3058R} tested two samples, observed with the OVRO 40m telescope at 15 GHz from 2008-2011, and found the same for their $\gamma$-ray selected AGN, but that BL Lacs were significantly more variable in their radio-selected sample.

Due to the association between Doppler factor and variability \citep{2001ApJ...549L..55T}, in a scenario where a lower synchrotron peak frequency corresponds with higher Doppler beaming, LSP BL Lacs would be expected to have the highest level of variability among the subclasses. \citealt{2014MNRAS.438.3058R} also reported a group of highly variable ISP BL Lacs, but additionally found that LSP BL Lacs were significantly more variable than HSP BL Lacs.

\begin{figure}
	\plotone{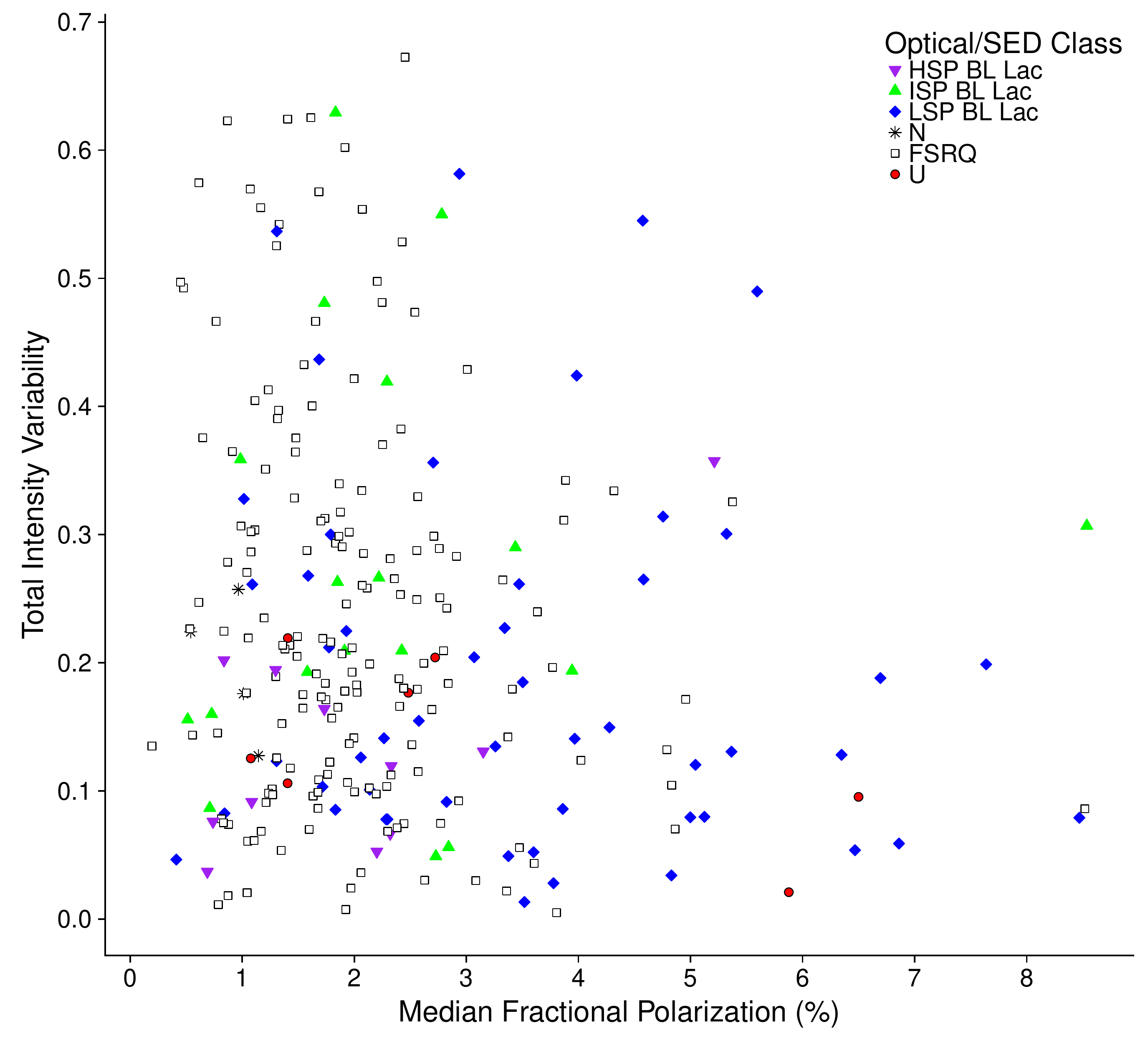}
  	\caption{\label{varimedms} Median fractional polarization $m_\mathrm{med}$ versus total intensity variability $I_\mathrm{var}$ for each AGN over time. Purple inverted triangles are HSP BL Lacs, green triangles are ISP BL Lacs, blue diamonds are LSP BL Lacs, black stars are NLSy1s, unfilled squares are FSRQs, and red circles have an unknown optical class or synchrotron peak frequency. A Kendall tau test of correlation yields $p=0.04$ for no correlation.}
\end{figure}

\begin{figure}
	\plotone{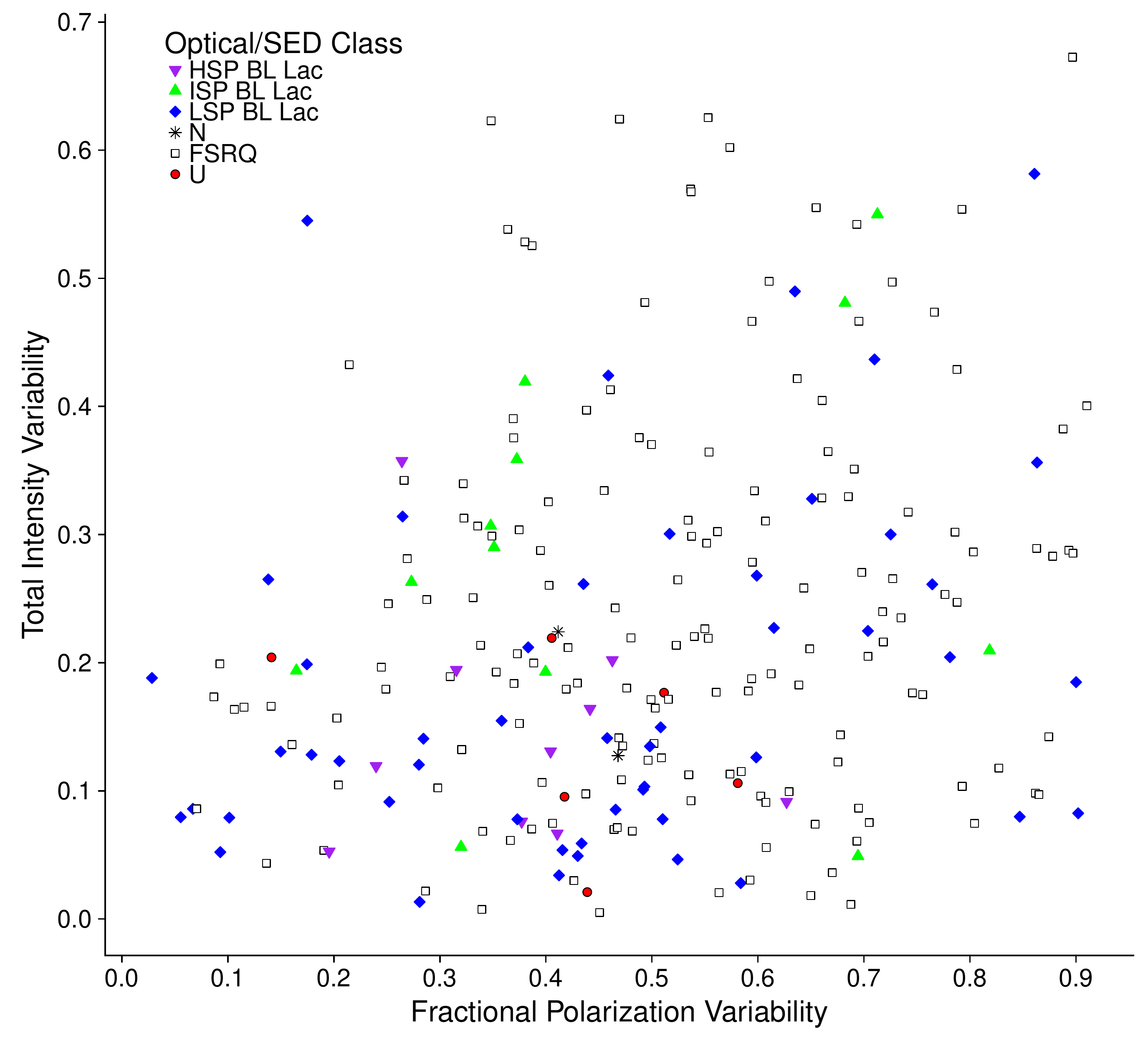}
  	\caption{\label{varivarm} Variability in fractional polarization $m_\mathrm{var}$ versus variability in total intensity $I_\mathrm{var}$ for each AGN over time. Purple inverted triangles are HSP BL Lacs, green triangles are ISP BL Lacs, blue diamonds are LSP BL Lacs, black stars are NLSy1s, unfilled squares are FSRQs, and red circles have an unknown optical class or synchrotron peak frequency. A Kendall tau test of correlation yields $p=0.003$ for no correlation.}
\end{figure}

Total intensity variability appears to be anti-correlated with fractional polarization in our sample, with a notable deficit of highly variable cores with high fractional polarization (Figure~\ref{varimedms}). We performed a Kendall tau non-parametric test of correlation, which yielded a $p=0.04$ probability for no correlation. This trend disappears if the BL Lac and FSRQ populations are tested separately, making it likely that the anti-correlation is simply due to the presence of low variability, highly polarized LSP BL Lacs. Values for $I_\mathrm{var}$ are generally lower than those of $m_\mathrm{var}$, indicating that the fractional polarization undergoes greater percentage change during the same time period. The two variability indices are correlated, with a tau test yielding $p=0.003$ for no correlation (see Figure~\ref{varivarm}). There is a lack of AGN cores which are variable in total intensity while remaining stable in fractional polarization. On the other hand, the fractional polarization can be highly variable even in cases where the total intensity shows little variability.

\subsection{EVPA Variability} \label{evpavar}

\begin{figure}
	\plotone{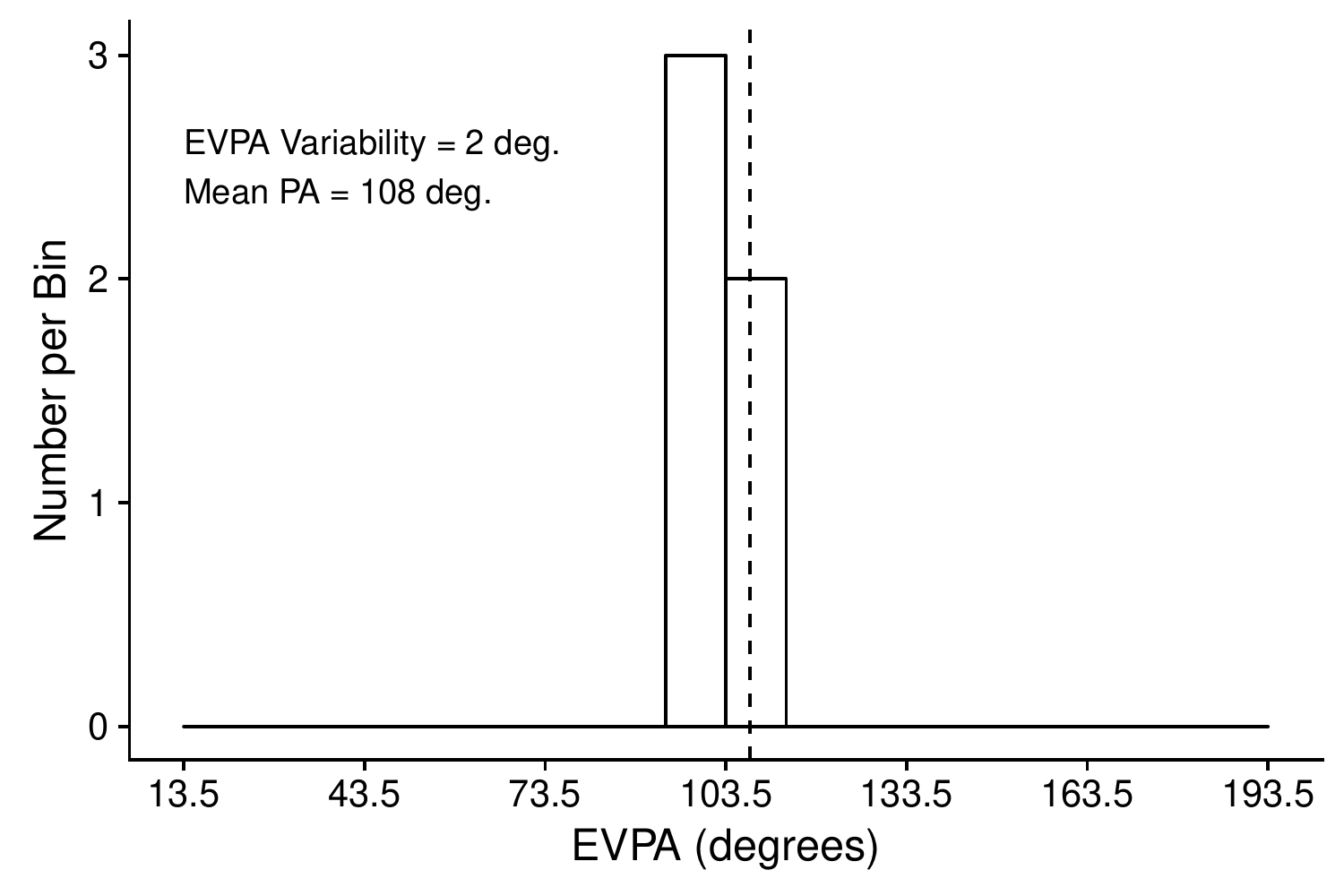}
    \caption{\label{lowsigma} The EVPAs used in calculating the standard deviation for GB6 J0929+5013. This AGN is an ISP BL Lac with low EVPA variability $EVPA_\mathrm{var}$. The circular mean of the jet PAs measured at the same epochs is displayed as a vertical dashed line.}
\end{figure}

\begin{figure}
	\plotone{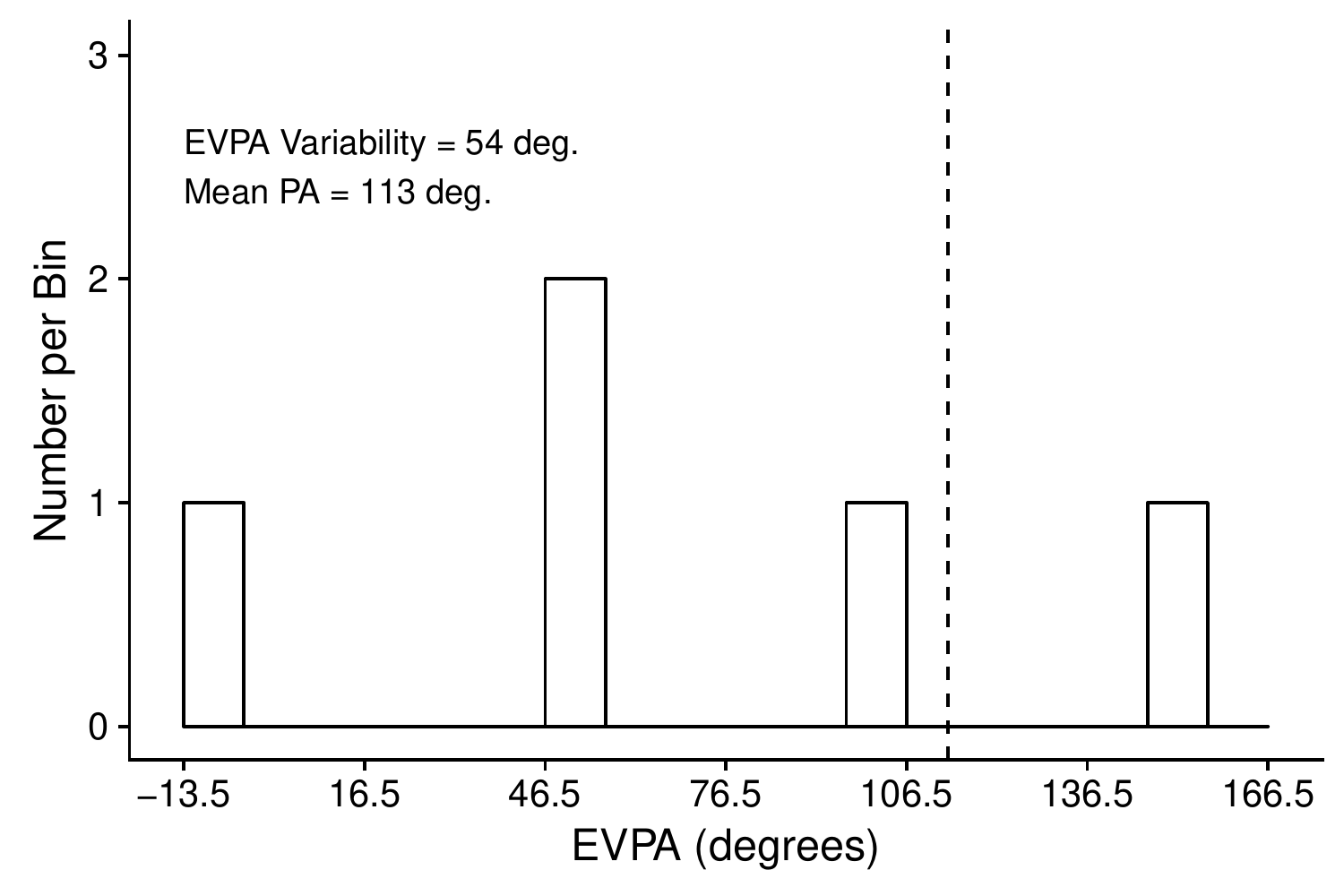}
    \caption{\label{highsigma} The EVPAs used in calculating the standard deviation for 3C 395. This AGN is an FSRQ with high EVPA variability $EVPA_\mathrm{var}$. The circular mean of the jet PAs measured at the same epochs is displayed as a vertical dashed line.}
\end{figure}

The characterization of EVPA variability is not straightforward due to the $180\arcdeg$ ambiguity in the measurements, and requires a special procedure.  We begin by adding or subtracting $180\arcdeg$ from all the EVPA measurements for a particular AGN such that they lie within the range of $0\arcdeg$--$180\arcdeg$. We then add a +$180\arcdeg$-shifted copy of each of these values, creating a total range of $0\arcdeg$--$360\arcdeg$. We bin this set of data using sequentially larger bins until at least one bin contains half the data. The midpoint of that bin is established as the mode of the EVPAs, and we then rotate the original (un-copied) set of EVPAs by the same amount (multiples of $180\arcdeg$) so that they lie within $\pm 90\arcdeg$ of this mode. Our $EVPA_\mathrm{var}$ statistic is defined as the standard deviation of these angles. In Figures~\ref{lowsigma} and~\ref{highsigma}, we show examples of cores with low and high EVPA variability, respectively.

\begin{figure}
	\plotone{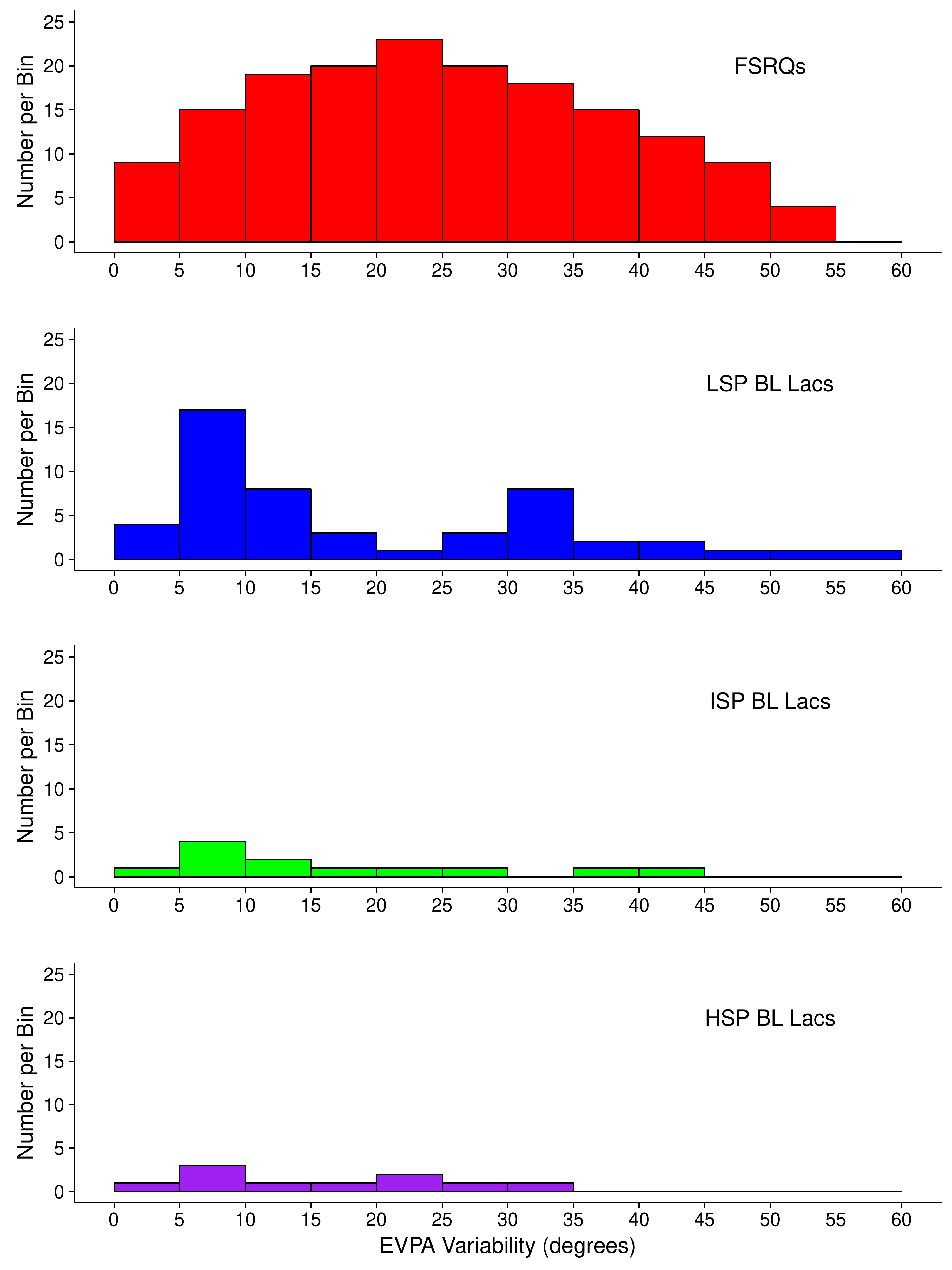}
    \caption{\label{sigmaevpa} Distributions of EVPA variability $EVPA_\mathrm{var}$, grouped by optical/synchrotron peak classification.}
\end{figure}

We plot the distributions of this statistic in Figure~\ref{sigmaevpa}. Examining them by optical class, the FSRQ variabilities follow an approximate normal distribution, but BL Lacs are skewed towards lower variability (KS test p-value of 0.0005 between FSRQs and LSP BL Lacs). In the optical, \citealt{2016MNRAS.463.3365A} reported an anti-correlation between synchrotron peak frequency and EVPA variability for a sample of 79 AGN; the mean of our BL Lac distributions become progressively lower with increasing synchrotron peak, but none of the subpopulations are significantly different, and a statistical test between synchrotron peak frequency and EVPA variability showed no significant correlation ($p=0.3$).

\begin{figure}
	\plotone{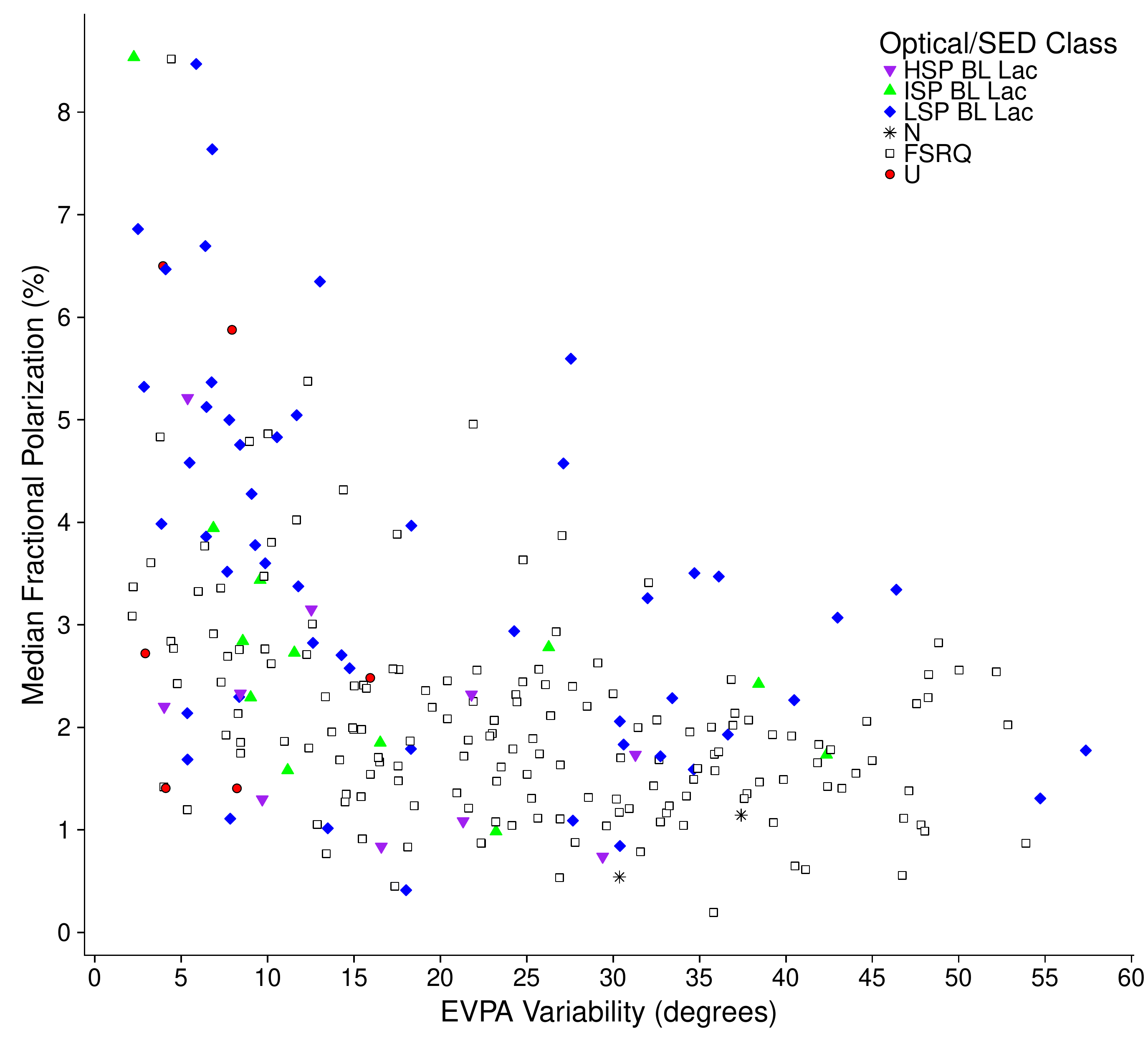}
  	\caption{\label{sigmamedms} EVPA variability $EVPA_\mathrm{var}$ versus median fractional polarization $m_\mathrm{med}$ for each AGN over time. Purple inverted triangles are HSP BL Lacs, green triangles are ISP BL Lacs, blue diamonds are LSP BL Lacs, black stars are NLSy1s, unfilled squares are FSRQs, and red circles have an unknown optical class or synchrotron peak frequency. A Kendall tau test of correlation yields $p=1.9\times 10^{-13}$ for no correlation.}
\end{figure}

\begin{figure}
	\plotone{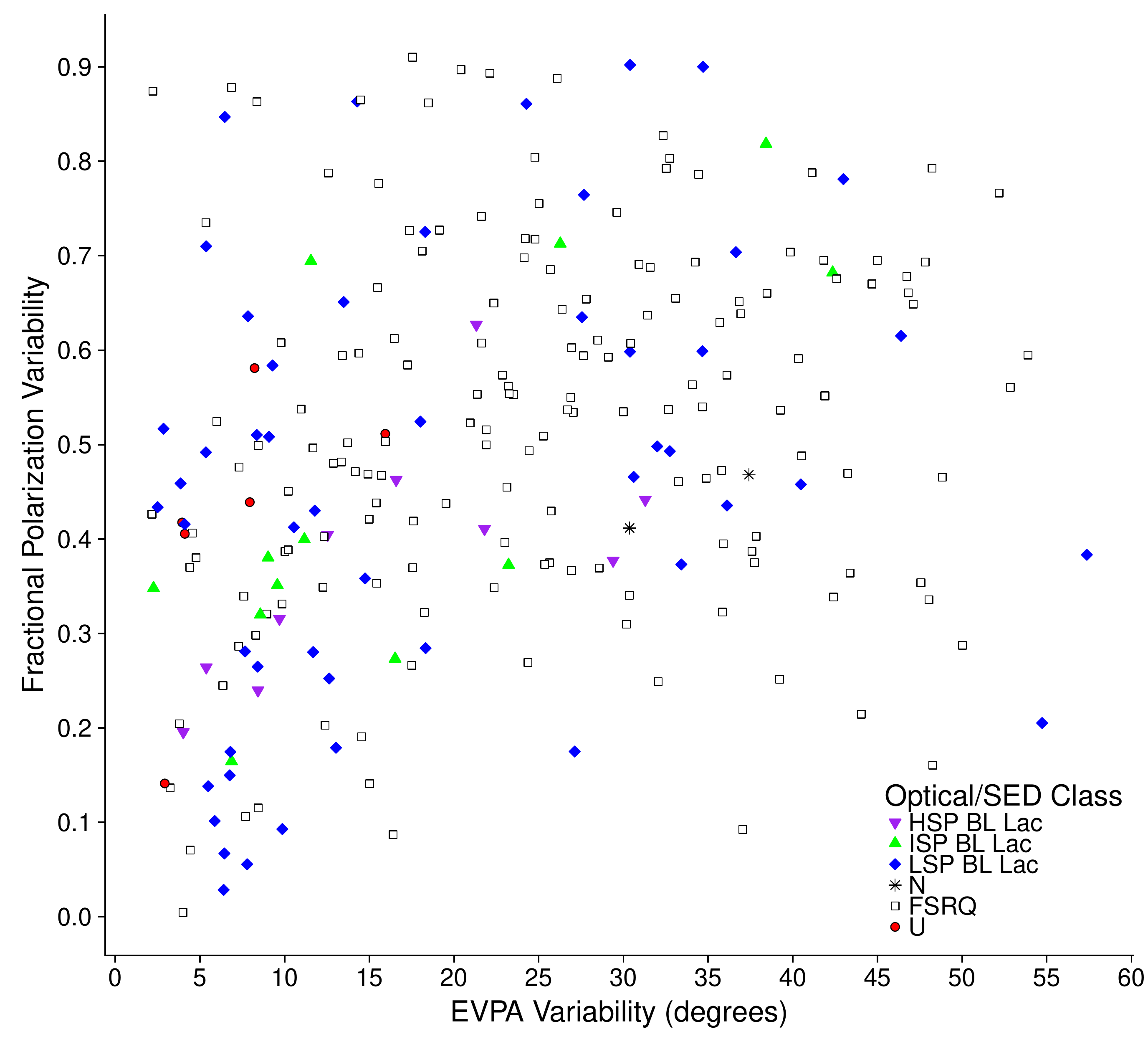}
  	\caption{\label{sigmavarms} EVPA variability $EVPA_\mathrm{var}$ versus fractional polarization variability $m_\mathrm{var}$ for each AGN over time. Purple inverted triangles are HSP BL Lacs, green triangles are ISP BL Lacs, blue diamonds are LSP BL Lacs, black stars are NLSy1s, unfilled squares are FSRQs, and red circles have an unknown optical class or synchrotron peak frequency. A Kendall tau test of correlation yields $p=4.1\times 10^{-7}$ for no correlation.}
\end{figure}

As was the case with the median EVPA-jet PA alignment, the EVPA variability is also anti-correlated with fractional polarization (Figure~\ref{sigmamedms}, $p=1.9\times 10^{-13}$) and positively correlated with fractional polarization variability (Figure~\ref{sigmavarms}, $p=4.1\times 10^{-7}$), i.e., AGN with preferred EVPA directions have greater fractional polarization which is more stable over time. These results are consistent with a strong, standing transverse shock at the base of the jet. When evidence for standings shocks is observed, they frequently appear to be near the cores of blazars in VLBI images \citep{2017ApJ...846...98J, MOJAVE_X}. It is possible that others may exist closer to the base of the jet, but are unseen due to the finite angular resolution of the VLBA at 15 GHz ($\simeq$0.5 mas) \citep{2016ApJ...817...96G}.

\begin{figure}
	\plotone{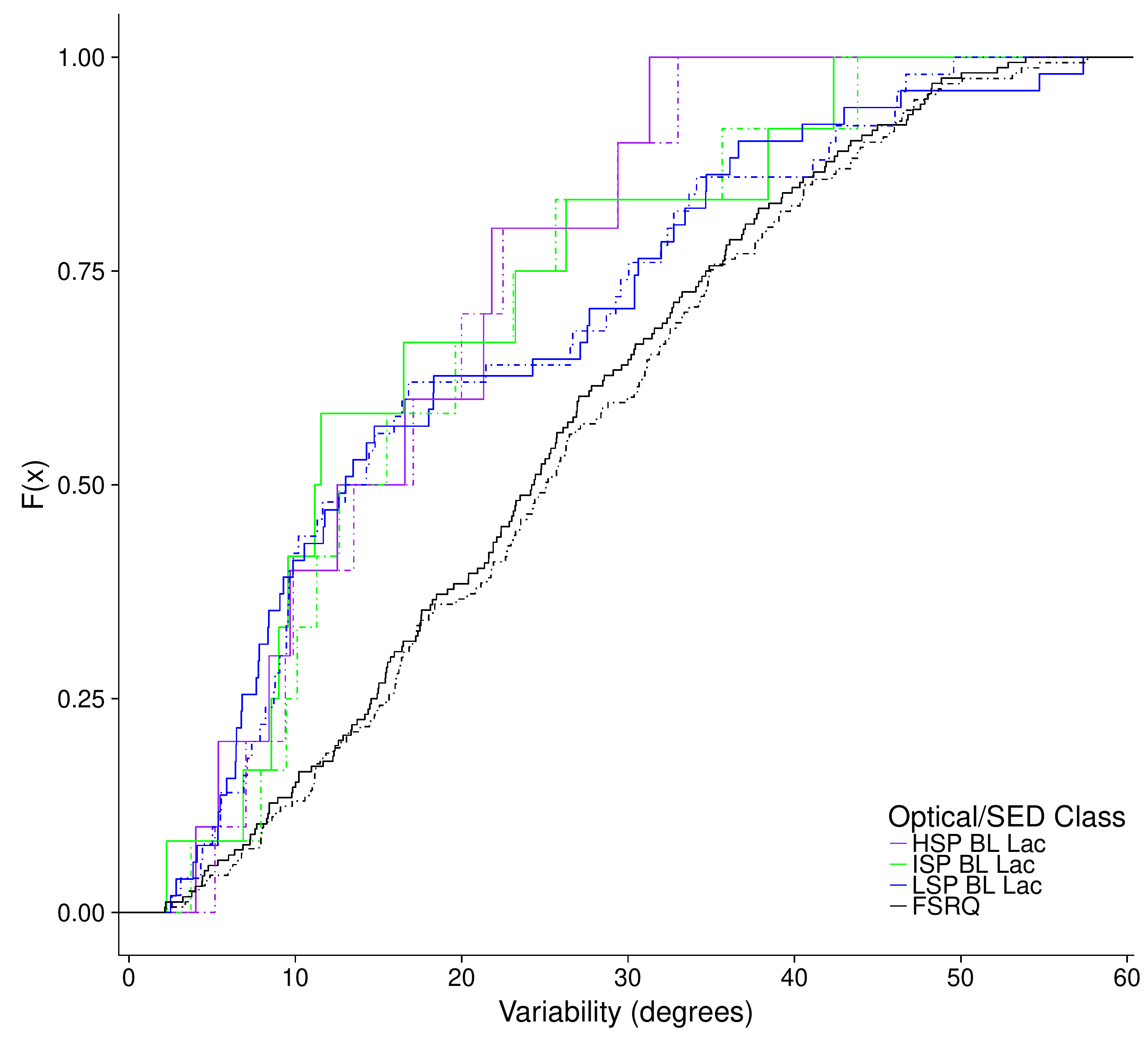}
    \caption{\label{sigmacdfs} Cumulative distribution functions of EVPA (solid lines) and EVPA-PA (dash-dotted lines) variability ($EVPA_\mathrm{var}$ and $(EVPA-PA)_\mathrm{var}$), grouped by optical/synchrotron peak classification. HSP BL Lacs are purple, ISP BL Lacs are green, LSP BL Lacs are blue, and FSRQs are black.}
\end{figure}

Our measure of EVPA variability does not include any information about the rotation of the EVPA with respect to the local jet direction. The jet PAs of MOJAVE AGN typically do not vary on the level of the core EVPAs, but there are some individual exceptions where systematic changes of up to a few degrees per year are seen \citep{2007AA...476L..17A, MOJAVE_X}. To address whether EVPA rotations coincide with rotations of the jet PA, we first subtracted the jet PA from the EVPA; essentially, taking all angles relative to the PA at each epoch rather than from the north direction on the sky. We then followed the method described for $EVPA_\mathrm{var}$ to obtain $(EVPA-PA)_\mathrm{var}$. We plot the empirical cumulative distribution functions (CDFs) of $(EVPA-PA)_\mathrm{var}$ and $EVPA_\mathrm{var}$ for the main optical classes in Figure~\ref{sigmacdfs}. For each of these classes, the CDF of $(EVPA-PA)_\mathrm{var}$ is similar to that of $EVPA_\mathrm{var}$, and the separation between classes is preserved, e.g. FSRQs are more variable than BL Lacs. As expected, taking into account the jet PA variations only slightly changes the overall EVPA variability statistics.

\subsection{EVPA Stability}

\begin{figure}
	\plotone{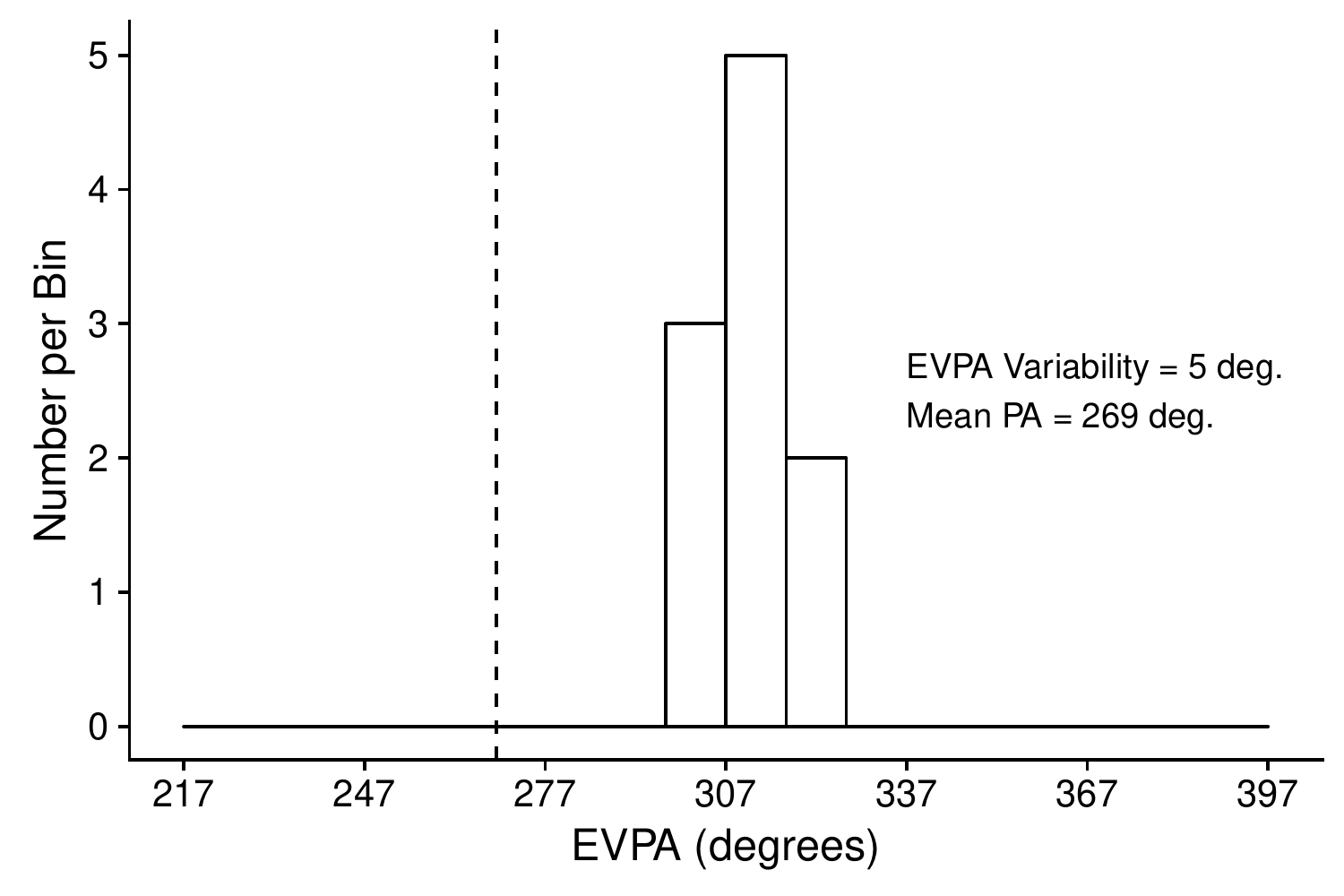}
    \caption{\label{ox161} The EVPAs used in calculating the standard deviation over all epochs for OX 161. This AGN has the lowest EVPA variability $EVPA_\mathrm{var}$ in the sample when all epochs are used in calculation. The circular mean of the jet PAs measured over all epochs is displayed as a vertical dashed line.}
\end{figure}

We find that many of the EVPAs are fairly stable---24\% of the AGN with available values have an $EVPA_\mathrm{var}$ of less than $10\arcdeg$. In contrast, \citealt{2003ApJ...586...33A} found "long-term stability" in only seven of their 62 sources at 14.5 GHz, potentially due to greater influence from components near the core which are not resolvable by a single dish telescope \citep{ALL99}. We have further investigated the possibility of AGN with stable EVPAs by also computing $EVPA_\mathrm{var}$ based on all epochs, rather than the relatively unbiased five epochs used in Section~\ref{evpavar} and the majority of this paper. We constrain this calculation to AGN with at least 10 measurements of EVPA; of the 171 AGN which meet this criterion, only 10 (listed in Table~\ref{evpastability}) have an $EVPA_\mathrm{var}$ of less than 10\arcdeg. Of these 10 AGN, eight have an EVPA mode value within 20\arcdeg of the mean jet PA.  This relationship agrees with the trends regarding EVPA-jet PA alignment and EVPA variability reported in the previous section. In Figure~\ref{ox161}, we show the histogram of EVPAs for OX 161, which is the least variable with an $EVPA_\mathrm{var}$ of 4.7\arcdeg. For all AGN with at least 10 measurements of EVPA, the median coverage time---from first epoch with EVPA to last epoch with EVPA---is 10.7 years. The percentage of AGN with stable EVPAs ($EVPA_\mathrm{var}$ under 10\arcdeg) is 6\%. This is a large decrease from the 24\% result calculated from 2.3 year periods, and is more consistent with the findings of \citealt{2003ApJ...586...33A}, which were based on $\sim$15 years of monitoring data. It is well-known that AGN are variable on long-term timescales in addition to their short-term variability \citep{2007A&A...469..899H}; the results presented here stress that long-term monitoring is essential to understanding the EVPA as well.

\begin{deluxetable*}{lcccccc}
\tablecolumns{7} 
\tabletypesize{\scriptsize} 
\tablewidth{0pt}  
\tablecaption{\label{evpastability} AGN with Stable Core EVPAs}  
\tablehead{ & & \colhead{Coverage Time} & & \colhead{$EVPA_\mathrm{var}$} & \colhead{EVPA Mode} & \colhead{Mean PA} \\
\colhead{Alias} & \colhead{Number of Epochs} & \colhead{(years)} & \colhead{Optical/SED Class} & \colhead{(deg)} & \colhead{(deg)} & \colhead{(deg)} \\
\colhead{(1)} & \colhead{(2)} & \colhead{(3)} & \colhead{(4)} & \colhead{(5)} & \colhead{(6)} & \colhead{(7)}} 
	\startdata 
	NRAO 005 & 18 & 9.7 & LSP B & 8.5 & 18.5 & 2 \\
	GB6 J0929+5013 & 11 & 5.1 & ISP B & 8.3 & 103.0 & 111 \\
	B2 1040+24A & 10 & 4.5 & Q & 8.5 & 96.0 & 87 \\
	OP 112 & 11 & 16.9 & LSP B & 6.4 & 35.0 & 40 \\
	OQ 240 & 11 & 7.1 & HSP B & 8.1 & 143.0 & 141 \\
	4C +14.60 & 11 & 14.6 & LSP B & 6.9 & 325.5 & 324 \\
	4C +56.27 & 26 & 14.6 & LSP B & 9.4 & 208.5 & 200 \\
	OV 591 & 11 & 6.1 & Q & 9.7 & 36.5 & 117 \\
	4C $-$02.81 & 16 & 15.3 & Q & 5.9 & 90.5 & 101 \\
	OX 161 & 10 & 7.8 & Q & 4.7 & 307.0 & 269 \\
	\enddata 
\tablecomments{Columns are as follows: (1) AGN name, (2) number of EVPA epochs used to calculate $EVPA_\mathrm{var}$, (3) time in years between first and last epoch with EVPA, (4) optical/SED peak classification where Q = FSRQ and B = BL Lac (all FSRQs are low-synchrotron-peaked), (5) EVPA variability in degrees calculated with all available EVPA epochs, (6) EVPA mode in degrees (see Section~\ref{evpavar}) rotated to within 180\arcdeg of the mean jet PA, (7) circular mean of the jet PA calculated with all available epochs.}
\end{deluxetable*} 

\subsection{Correlations with Luminosity}

\begin{figure}
	\plotone{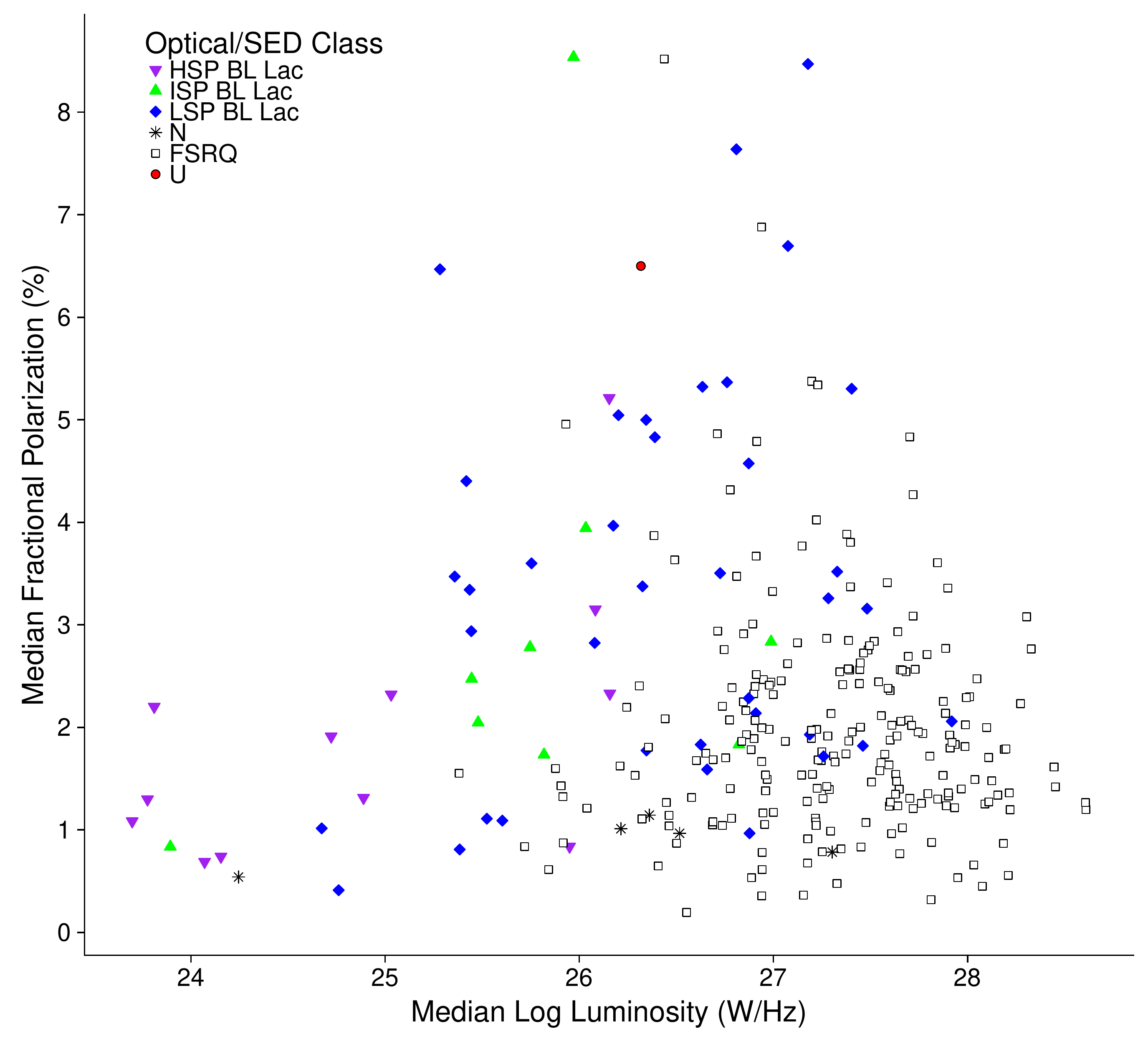}
  	\caption{\label{luminms} Median luminosity $L_\mathrm{med}$ versus median fractional polarization $m_\mathrm{med}$ for each AGN over time. Purple inverted triangles are HSP BL Lacs, green triangles are ISP BL Lacs, blue diamonds are LSP BL Lacs, black stars are NLSy1s, unfilled squares are FSRQs, and red circles have an unknown optical class or synchrotron peak frequency. There is no significant correlation. A Kendall tau test of correlation yields $p=0.007$ for no correlation if FSRQs are excluded, however.}
\end{figure}

We have calculated the median core luminosity of each AGN with known redshift, assuming a flat spectral index, as: \begin{equation} L_\mathrm{med} = \frac{4 \pi d_\mathrm{L}^{2} I_\mathrm{med}}{(1+z)}, \end{equation} where $d_\mathrm{L}^{2}$ is the luminosity distance and $I_\mathrm{med}$ is the median core intensity. The latter is approximately equal to the core flux density, since the cores are typically unresolved. In Figure~\ref{luminms}, we plot the median luminosity versus the median fractional polarization for all available cores. The lowest core luminosities are associated with low core fractional polarization, and consist primarily of HSP BL Lacs. Given that the highly compact core emission is most likely Doppler boosted, this lends further support to the scenario described in \citealt{2011ApJ...742...27L}, in which HSP BL Lacs are less beamed and also less polarized due to lower Doppler factors; this trend would also be expected based on the model of \citealt{2016MNRAS.463.3365A}. However, we find no relationship between our EVPA-based statistics and the median luminosity, which suggests that the low polarization is unrelated to inherent magnetic field properties. 

\subsection{Narrow Line Seyfert 1s}

Five of the AGN in our sample are classified as narrow line Seyfert 1 galaxies (NLSy1). All of them have a relatively low median fractional polarization, with the highest reaching only 1.14\%. They also have EVPAs misaligned with the jet PA, with $|EVPA-PA|_\mathrm{med}$ ranging from $44\arcdeg$ to $80\arcdeg$. For four of these five, we were able to calculate the variability in total intensity. These values are also relatively low compared to blazars, with the highest $I_\mathrm{var}$ equal to 0.26 (the mean FSRQ value is 0.24). Due to the number of measurements qualified as upper limits, the fractional polarization variability and EVPA variability could only be determined for two of the NLSy1s.

\subsection{Correlations with $\gamma$-ray Emission}

In this section, we parallel the analysis of Sections~\ref{fracpol} through~\ref{evpavar} with a basis on AGN detection status by \(Fermi's\) Large Area Telescope (LAT). Because of the lower energy cutoff of the LAT instrument, it more readily detects AGN with higher synchrotron peak frequencies. Conversely, AGN selected on the basis of bright radio emission prefer lower peaked sources, representing a separate population \citep{2011ApJ...742...27L}. The \(Fermi\)-LAT AGN catalog is almost entirely dominated by blazars, which suggests that the brightest $\gamma$-ray AGN have high Doppler beaming factors \citep{2015ApJS..218...23A}. Therefore, an AGN sample selected on the basis of low-frequency radio emission (e.g., the 3CR survey \citealt{1962MNRAS.125...75B}), in which lobe-dominant objects are common, will have fewer LAT detections since it is non-orientation biased. In order to limit our analysis to radio-selected blazars, we only examine the LAT detection statistics of AGN included in the MOJAVE 1.5 Jy VLBA flux-density limited sample, and exclude the radio galaxies, NLSy1s, and optically unidentified sources. We also exclude the galactic plane region ($\mid$b$\mid$ $>10\arcdeg$) due to the greater difficulties of identifying reliable \(Fermi\) associations \citep{2015ApJS..218...23A}. We ultimately consider 155 AGN with optical class as follows: 97 FSRQs and 24 BL Lacs associated with a LAT detection; and 33 FSRQs and 1 BL Lac with no \(Fermi\) third source catalog \citep{2015ApJS..218...23A} association.

\begin{figure}
	\plotone{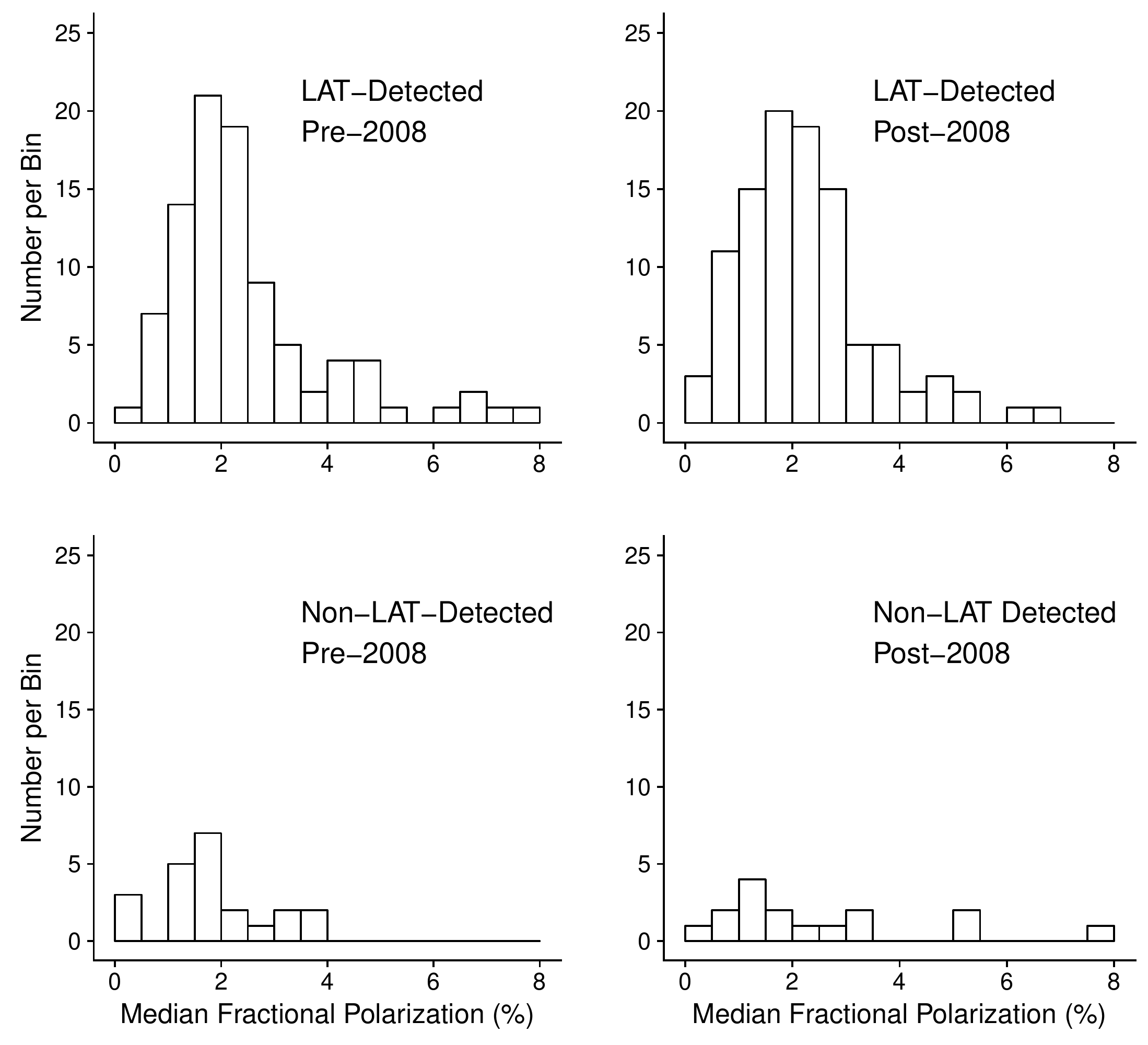}
    \caption{\label{latmedms} Distributions of median fractional polarization $m_\mathrm{med}$, calculated separately using all epochs pre-August 4th, 2008 (left) and all epochs after (right) based on the start of {\it Fermi}-LAT observations. AGN are grouped as LAT-detected (top) or non-LAT-detected (bottom).}
\end{figure}

We begin by revisiting the conclusions of \citealt{2010IJMPD..19..943H}, an early study which investigated $\gamma$-ray connections with radio polarization using a similar subset of the original 1.5 Jy Sample (a total of 123 AGN) and the \(Fermi\) LAT Bright Source List (0FGL; \citealt{2009ApJ...700..597A}). In \citealt{2010IJMPD..19..943H}, 0FGL AGN (those detected in the first three months of the \(Fermi\) mission) were found to have greater polarized flux than non-0FGL AGN during the year post-\(Fermi\) launch. The median fractional polarization of 0FGL AGN was also higher than non-0FGL AGN in the same year, but no difference was found between medians calculated from the six years pre-launch. We have similarly divided our observations into pre- and post-launch periods (before and after August 2008)  and calculated the median fractional polarization for each AGN separately within the time frames. We plot the results in Figure~\ref{latmedms}. In contrast to \citealt{2010IJMPD..19..943H}, we do not find LAT-detected AGN to have a higher median fractional polarization during the LAT observing era. We also find no significant differences between the LAT-detected and non-LAT-detected AGN in either era.

However, non-LAT-detected AGN more often lack a measurable median fractional polarization due to higher numbers of upper limit measurements---12\% of non-LAT-detected AGN have unknown $m_\mathrm{med}$ in the pre-LAT era, and 20\% in the post-LAT era. By comparison, 4\% and 3\% of LAT-detected AGN have unknown $m_\mathrm{med}$ in the pre- and post-LAT eras, respectively. This agrees with the VLBA 5 GHz findings of \citealt{2012ApJ...744..177L}, which claimed non-LAT-detected AGN were more often unpolarized, but when their measurements were polarized enough to be qualified as detections, they had comparable fractional polarization to LAT-detected AGN. We investigated further by dispensing with the use of $m_\mathrm{med}$ and directly comparing the fractional polarization values, including censored points, of the five epochs per AGN described in Section~\ref{stattests}. This was done without dividing the data into pre- and post-LAT eras. Our statistical test did not reveal significant differences ($p=0.15$). In Section~\ref{fracpol}, we discussed differing fractional polarizations among optical classes, and the LAT-detected set of AGN has a far greater ratio of BL Lacs to FSRQs. We note that the significance of the preceding results, or lack thereof, does not change when the analysis is limited to only FSRQs.

\begin{figure}
	\plotone{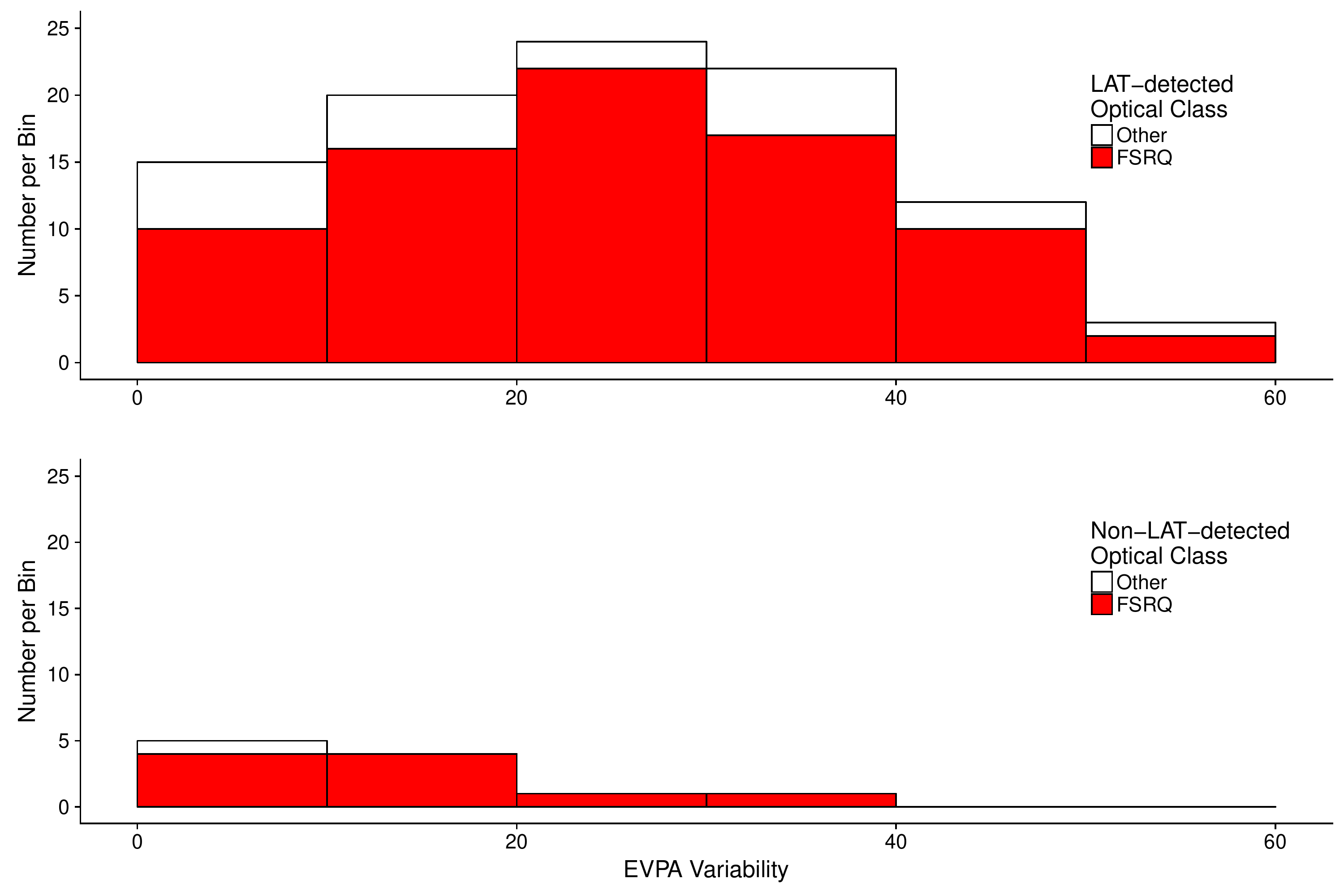}
    \caption{\label{latsigmas} Distributions of EVPA variability $EVPA_\mathrm{var}$. AGN are grouped as LAT-detected (top) or non-LAT-detected (bottom). Unfilled bins represent the entire sample and red bins represent FSRQs.}
\end{figure}

Previous research has suggested that LAT-detected AGN are more variable in both total intensity \citep{2014MNRAS.438.3058R} and fractional polarization \citep{2010IJMPD..19..943H}. In our sample, both LAT-detected variability distributions have greater means, but we cannot confirm significant differences (KS test $p=0.13$ for total intensity variability and $p=0.13$ for fractional polarization variability as well). While we find no difference in median EVPA-jet PA offset based on LAT detection status, we do find a significant difference in EVPA variability. In Figure~\ref{latsigmas}, we show that non-LAT-detected AGN are skewed toward low values of $EVPA_\mathrm{var}$, and a KS test results in a p-value of 0.005 for the same parent distribution ($p=0.008$ if only FSRQs are compared).

Higher EVPA variability in LAT-detected AGN could be explained by a connection between $\gamma$-ray flaring and EVPA rotation. Optical EVPA rotation events have only been observed in $\gamma$-ray-loud AGN \citep{2018MNRAS.474.1296B}. The relationship between radio EVPA events and $\gamma$-ray flaring is less certain; \citealt{2001ApJ...556..738J} found a statistical link between  periods of high $\gamma$-ray activity and the ejection of radio knots, partially through analysis of the radio polarization. Additionally, the radio polarization behavior of multiple AGN have been modeled at the time of $\gamma$-ray flares with data from the UMRAO \citep{2014ApJ...791...53A, 2016Galax...4...35A}. These shock-based models, however, also predict increases in fractional polarization and radio flux density during $\gamma$-ray flares. Greater variability in fractional polarization and total intensity might then be expected for LAT-detected AGN, which is not seen to a significant degree in our data.

\section{Summary}

We have investigated the linear polarization properties of a sample of 387 parsec-scale AGN jet cores, using 15 GHz VLBA data. In the first paper in the series, it was found that BL Lacs in general are more polarized than FSRQs and have EVPAs which are better aligned with the local jet direction \citep{MOJAVE_I}. We have confirmed these findings and expanded on them with a larger AGN sample representing a broader range of synchrotron peak frequencies. With new multi-epoch measurements, we have also explored a variety of statistics related to AGN variability. Our conclusions are as follows:
\begin{enumerate}

\item Although HSP BL Lacs and FSRQs have similar median fractional polarization, LSP BL Lacs are significantly more polarized. Radio galaxies have too many censored data points for calculation of a median, which suggests that they have relatively low fractional polarization. 

\item BL Lacs have EVPAs that tend towards alignment with the jet PA, while FSRQs are skewed towards misalignment. BL Lac populations have similar median EVPA-jet PA alignment regardless of synchrotron peak. The most polarized cores have EVPAs almost parallel with the jet PA.

\item HSP BL Lacs are less variable in fractional polarization than FSRQs, but we find no significant differences among other optical/SED peak classes. The least variable AGN cores tend to have EVPAs aligned with the local jet direction.

\item ISP BL Lacs appear the most variable in total intensity; the HSP and LSP BL Lac distributions are similar, and both less variable than FSRQs. Fractional polarization is more variable than total intensity, and the two variabilities are positively correlated.

\item The EVPAs of BL Lacs are less variable than the EVPAs of FSRQs, and they do not appear dependent on synchrotron peak frequency. AGN cores with low EVPA variability are more likely to have high fractional polarization and low variability in fractional polarization. Our results show little change when the jet PA is subtracted from the EVPA before calculation of the variability; in other words, the angle difference between the jet PA and EVPA is similar in variability to the EVPA alone.

\item Of the five NLSy1s in our sample, all have low fractional polarization compared to blazars. Additionally, their EVPAs are not aligned with the jet PA.

\item AGN detected at $\gamma$-ray energies by \(Fermi\)-LAT are not significantly more fractionally polarized than non-detected AGN. However, their EVPAs are significantly more variable.

\end{enumerate}Overall, we believe these results are indicative of inherent differences between BL Lacs and FSRQs, perhaps in shock strength and geometry; by contrast, differences in polarization based on synchrotron peak and $\gamma$-ray detection can generally be explained by Doppler boosting. Analysis of the downstream jet polarization properties will be presented in a future paper in this series.

The MOJAVE project was supported by NASA-Fermi grants NNX08AV67G, NNX12A087G, and NNX15AU76G. MFA was supported in part by NASA-Fermi GI grants NNX09AU16G, NNX10AP16G, NNX11AO13G,\linebreak NNX13AP18G and NSF grant AST-0607523. YYK and ABP were supported by the Russian Foundation for Basic Research (project 17-02-00197), the Basic Research Program P-28 of the Presidium of the Russian Academy of Sciences, as well as by the government of the Russian Federation (agreement 05.Y09.21.0018). TS was supported by the Academy of Finland projects 274477 and 284495. The Long Baseline Observatory and the National Radio Astronomy Observatory are facilities of the National Science Foundation operated under cooperative agreement by Associated Universities, Inc.  This work made use of the Swinburne University of Technology software correlator \citep{2011PASP..123..275D}, developed as part of the Australian Major National Research Facilities  Programme and  operated under licence.

\software{AIPS \citep{2003ASSL..285..109G}, Difmap \citep{difmap}, R core package \citep{Rcore}, R NADA package \citep{RNADA}}

\appendix
{\centering{NOTES ON INDIVIDUAL SOURCES}}
\medskip

\par $0640+090$ (PMN J0643$+$0857): The exact core location in this low galactic latitude ($b = 2.3\arcdeg$) quasar is uncertain. We assigned the core to the most compact feature. There are three jet features to the west, and one feature to the east of the putative core.

\par 1118$+$073 (MG1 J112039$+$0704): We assigned the core location in this quasar to the most northeastern jet feature, which is not the most compact feature in the jet. 

\par 1148$-$001 (4C $-$00.47):  We assigned the core location to the most compact feature in this quasar. At some epochs, a weak jet feature appears to be present to the northeast, whereas the bulk of the jet emission lies to the southwest. 

\par 1435$+$638 (VIPS 0792):  We assigned the core location to the most northwestern feature in this quasar, which is not the most compact feature in the jet. 

\par 2234$+$282 (CTD 135):  This BL Lac jet has numerous bright features within 1 mas of each other, making the core identification uncertain. The most southwestern feature, which we assign as the core,  is the most compact feature in the MOJAVE epochs after 2009, and is the most compact feature in an unpublished 43 GHz VLBA image by T. An (2017, private communication).

\bibliographystyle{yahapj}
\bibliography{hodge}

\end{document}